\documentclass[final,3p,10pt]{elsarticle}

\usepackage{graphicx}
\usepackage{xcolor}
\usepackage{soul}
\usepackage{algorithm}
\usepackage{algpseudocode}
\usepackage{siunitx}
\usepackage{amssymb}
\usepackage{amsthm}
\usepackage{amsfonts}
\usepackage{mathtools}
\usepackage{multirow}
\usepackage{float}
\usepackage{hyperref}

\usepackage{lipsum}
\newcommand\blfootnote[1]{%
  \begingroup
  \renewcommand\thefootnote{}\footnote{#1}%
  \addtocounter{footnote}{-1}%
  \endgroup
}

\journal{Composites Part A: Applied Science and Manufacturing}
\everymath{\displaystyle}

\begin{document}

\blfootnote{This manuscript version is made available under the CC-BY-NC-ND 4.0 license.}
\blfootnote{DOI:~\url{https://doi.org/10.1016/j.compositesa.2022.107424}}
\blfootnote{DOI arXiv:~\url{https://doi.org/10.48550/arXiv.2301.05552}}

\begin{frontmatter}


\title{Application of the partial Dirichlet-Neumann contact algorithm to simulate low-velocity impact events on composite structures}

\author[bsc]{G. Guillamet\corref{cor1}}\ead{gerard.guillamet@bsc.es}
\author[bsc,udg,icl]{A. Quintanas-Corominas}
\author[bsc]{M. Rivero}
\author[bsc]{G. Houzeaux}
\author[bsc]{M. V\'{a}zquez}
\author[udg]{A. Turon}
\cortext[cor1]{Corresponding author}
\address[bsc]{Barcelona Supercomputing Center (BSC), Plaça Eusebi Güell, 1-3, Barcelona, 08034, Catalonia, Spain}
\address[udg]{AMADE, Universitat de Girona, Av. Universitat de Girona 4, Girona, 17003, Catalonia, Spain}
\address[icl]{Department of Civil and Environmental Engineering, Imperial College London, London, SW7 2AZ, UK}
                      
\begin{abstract}

Impact simulations for damage resistance analysis are computationally intensive due to contact algorithms and advanced damage models. Both methods, which are the main ingredients in an impact event, require refined meshes at the contact zone to obtain accurate predictions of the contact force and damage onset and propagation through the material. This work presents the application of the partial Dirichlet-Neumann contact algorithm to simulate low-velocity impact problems on composite structures using High-Performance Computing. This algorithm is devised for parallel finite element codes running on supercomputers, and it is extended to explicit time integration schemes to solve impact problems including damage. The proposed framework is validated with a standard test for damage resistance on fiber-reinforced polymer matrix composites. Moreover, the parallel performance of the proposed algorithm has been evaluated in a mesh of 74M of elements running with 2400 processors.

\end{abstract}




\end{frontmatter}


\section{Introduction}
\label{sec:intro}

Impacts by foreign objects against any part of the aircraft are a major concern for the aerospace industry because they may compromise the structural integrity of the aircraft. Impact events can be classified into three main categories: low, high (including ballistics), and hyper-high velocity impacts \citep{Abrate1994}. During the impact, the energy by the foreign object (projectile) is transferred to the target (structure), and consequently, the material can be damaged. Concretely, low-velocity impact events on composite materials (e.g., tool drops during maintenance or manufacturing) can drastically reduce the residual strength of the part even for case scenarios of Barely Visible Impact Damage (BVID). Therefore, designing composite structures with damage resistance cannot be avoided. Moreover, extensive experimental campaigns particularly focused on the investigation and evaluation of damage resistance of a specific material may be prohibitive by the industry in terms of costs.

Thus, virtual testing of impact events is of great interest as mathematical models and technology advance. However, solving the physics behind this problem, particularly from the material point of view, is still one of the most complex and challenging problems today. A review of existing software for composite impact modeling focused on low-velocity events is conducted by \citet{Nguyen2005}. In this review, the constitutive damage models play an essential role apart from the methods such as the contact algorithm or temporal integration scheme. Most of them can capture the trends and peak forces reasonably well. The research community has put and continues to put a lot of effort into developing reliable constitutive damage models for composites. Remarkable progress has been made on specific methodologies and constitutive damage models for predicting the damage resistance and damage tolerance of composite structures \citep{Lopes2009,Bouvet2009,Gonzalez2012,Tan2015,Lopes2016b,Soto2018b,Soto2018,Furtado2019}. However, in terms of computational performance, the resolution of an impact problem, including different sources of damage, is still computationally demanding. The use of sophisticated contact algorithms and advanced damage models require refined element meshes to accurately predict the onset and propagation of the damage in such materials.       

The most commonly used contact algorithms for the resolution of an impact problem are the Penalty methods \citep{Har2003}, Classical Lagrange multipliers \cite{Bathe1985,Gallego1989a} or the Augmented Lagrange multipliers. The latter is often chosen to solve the contact inequality constraints, see \cite{Weyler2012}. However, the parallel aspects of these traditional contact algorithms are not trivial, and to the authors' knowledge, little effort has been invested in the parallel aspects of such algorithms and their scalability in supercomputers. Some research works dealing with the parallel aspects of contact algorithms are \cite{Malone1994a,Malone1994b,Har2003}.

A completely different approach to the previous algorithms is the method of partial Dirichlet-Neumann (PDN) conditions. The contact is tackled as a coupled problem, in which the contacting bodies are treated separately, in a staggered way. The coupling is performed through the exchange of boundary conditions at the contact interface following a Gauss-Seidel strategy. The pioneering works using this approach are conducted by \citet{Krause2002} and \citet{YastrebovPhd2011}, showing the capabilities of solving nonlinear contact problems. To the authors' knowledge, one of the first applications of this method for explicit dynamics is made by \citet{Lapeer2019}, where the PDN method was used to simulate natural childbirth using explicit dynamics and executed in a hybrid system with Central Processing Units (CPU) and Graphics Processing Unit (GPU) architectures. However, little attention is dedicated to the computational performance and the parallel aspects of dealing with large-scale models. More recently, this method has been adapted and implemented in parallel in the \verb|Alya| multiphysics code \citep{Vazquez2016a} by \citet{Rivero2018PhD} and published by the authors in \citep{Guillamet2022a}. The mathematical and the parallel aspects are described in detail in these works, demonstrating the benefits of the PDN contact algorithm in High-Performance Computing (HPC) systems.

In this paper, we present the application of the aforementioned method proposed by the authors in \citep{Rivero2018PhD,Guillamet2022a} for the resolution of low-velocity impact problems for composite materials. Existing time integration schemes and constitutive models from the literature have also been adapted and implemented within a parallel framework. So the main contribution of the present paper is focused on the extension of the PDN contact algorithm for explicit time integration schemes and its use in HPC systems involving impact events in the field of composite materials. Additionally, a new mesh multiplication algorithm is presented to deal with cohesive elements and element technologies such as continuum shell elements.

The content of this paper is structured as follows. Firstly, the methods for the resolution of low-velocity impact events on composite materials including damage are explained with a strong emphasis on the contact algorithm and its implementation in parallel codes based on the finite element method. Then, the algorithm is validated through three benchmark tests. The first one consists of a quasi-static indentation test to verify that the contact pressure is well captured by using implicit and explicit time integration schemes. The second and the third examples correspond to a low-velocity impact on a composite plate following the ASTM International standard to measure the damage resistance of fiber-reinforced polymers. These last examples, use different material systems which are quite used by the aerospace industry, the T800S/M21 and the AS4/8552 carbon/epoxy systems. The numerical predictions obtained are correlated with experiments, and the computational performance is analyzed and discussed for the coupon made of AS4/8552 material. Finally, the conclusions of this work are commented on together with future work to improve the simulation of impact events including damage. 

\section{Modeling framework for low velocity impact events using High-performance systems}
\label{sec:framework}

This section describes the modeling framework and the application of the partial Dirichlet-Neumann (PDN) contact algorithm for the simulation of impact events. Particular emphasis is put on the extension of such contact algorithm for explicit dynamic analysis. All the methods presented here are implemented in the \verb|Alya| multiphysics code \citep{Vazquez2016a} based on the Finite Element Method (FEM). This parallel code is based on high-performance programming techniques for distributed and shared memory supercomputers. Moreover, the methods are programmed using the total Lagrangian formulation, where stresses and strains are measured with respect to the original configuration. The Green strain measure and the 2nd Piola-Kirchoff stress are used, and we follow the notation from Belytschko et al. \cite{Belytschko2014} throughout the paper. As an impact event is a complex and computationally demanding engineering problem is very attractive to be solved using High-Performance Computing. It is worth highlighting that all the methods described here can be implemented in other parallel FEM codes.

\subsection{Partial Dirichlet-Neumann contact algorithm}\label{sec:contalgo}

The low-velocity impact event proposed in this paper can be assumed as a non-linear contact problem, where the striker is considered as a rigid body and the plate as a deformable body. Let's assume that both body instances are of arbitrary shape, and we do not consider friction. Therefore, this contact problem can be written as a boundary value problem, see Yastrebov \citep{Yastrebov2013}, which includes the Hertz-Signorini-Moreau law for normal contact. So the balance of momentum and the contact conditions can be written as follows:

\begin{equation}\label{eq:bvp}
\begin{split}
& \nabla \cdot \underline{\underline{\boldsymbol{\sigma}}} + \underline{\boldsymbol{f}}_v = 0 \quad \textrm{in} \,\, \Omega\\
& \underline{\underline{\boldsymbol{\sigma}}} \cdot \underline{\boldsymbol{n}} = \underline{\boldsymbol{\sigma}}_0 \quad \textrm{on} \,\, \Gamma_N\\
& \underline{\boldsymbol{u}} = \underline{\boldsymbol{u}}_0 \quad \textrm{on} \,\, \Gamma_D \\
& g \geq 0, \,\, \sigma_n \leq 0, \,\, \sigma_n\,g = 0, \,\, \underline{\boldsymbol{\sigma}}_t = 0  \quad \textrm{on} \,\, \Gamma_C
\end{split}
\end{equation}

\noindent being $\underline{\underline{\boldsymbol{\sigma}}}$ the Cauchy stress tensor, $\underline{\boldsymbol{f}}_v$ a vector of volumetric forces, $\underline{\boldsymbol{\sigma}}_0$ a set of prescribed tractions on the Neumann boundary, $\Gamma_N$; $\underline{\boldsymbol{u}}_0$ a set of prescribed displacements on the Dirichlet boundary, $\Gamma_D$. Over the contact boundary, $\Gamma_C$, we have imposed the following conditions: $g$ represents the gap between contacting bodies, $\sigma_n$ is the normal contact pressure, and $\underline{\boldsymbol{\sigma}}_t$ is the tangential stress. The tangential stress equal to zero ($\underline{\boldsymbol{\sigma}}_t = 0$) in Eq.~\eqref{eq:bvp} characterizes a frictionless contact case.

In order to satisfy the conditions in Eq.~\eqref{eq:bvp}, the present paper uses the partial Dirichlet-Neumann contact algorithm proposed by Rivero \citep{Rivero2018PhD,Guillamet2022a} which is based on the work from Yastrebov \cite{Yastrebov2013}. In the works mentioned above, the method was applied for implicit time integration schemes, while in the present work, the algorithm is extended to explicit schemes. The main benefits of the PDN contact algorithm to typical Penalty or Lagrange Multipliers methods are the following: (i) the size of the problem does not increase due to the Lagrange multipliers methods as unknowns (ii) no restriction with respect to the mesh partitioner due to the use of contact elements (iii) absence of contact tangent matrices (implicit schemes) and residual contact force vectors and (iv) easy to be parallelized as it can be treated as a solid-to-solid coupling using existing methods for multiphysics applications such as the Gauss-Seidel scheme. 

\begin{figure*}[htp!]
\begin{center}
\includegraphics[width=\textwidth]{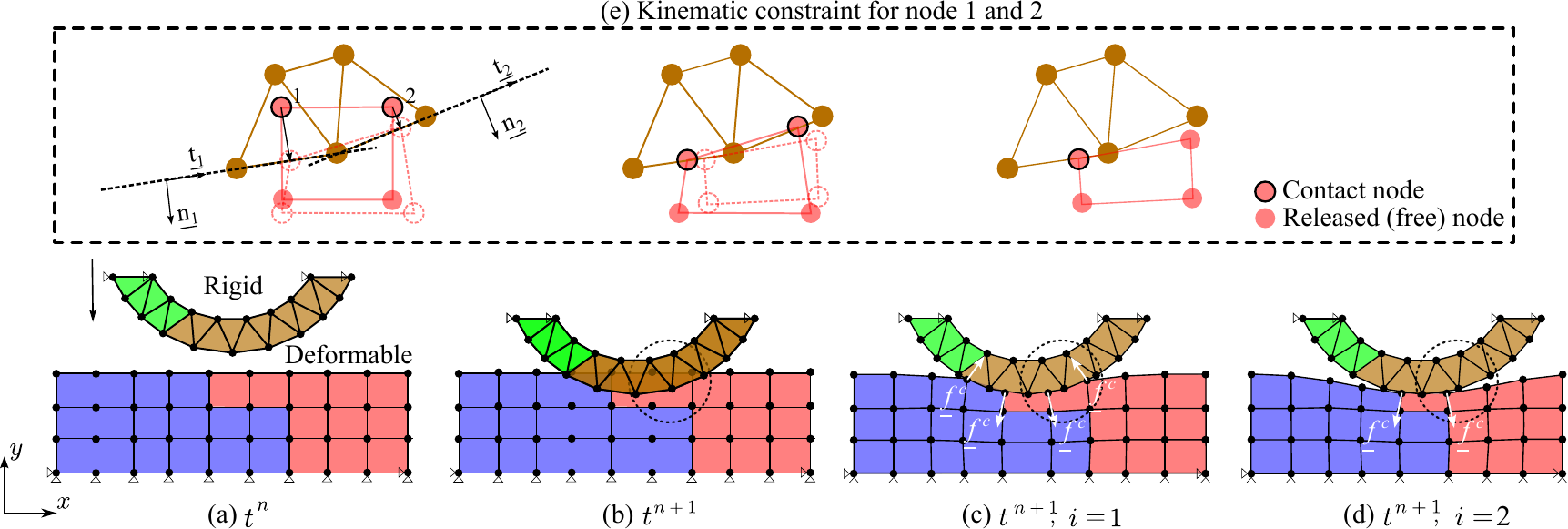} 
\caption{Iterative process of the parallel PDN contact algorithm. (a) Interaction; (b) Overlapping; (c) Dirichlet boundary conditions (projections) (d) Released nodes and equilibrium. (e) Kinematic constraint for node 1 and 2. The reader is referred to the web version of this paper for the color representation of this figure.}
\label{fig:workflow}
\end{center}
\end{figure*}

The iterative process of the PDN contact algorithm in a frictionless problem is shown in Fig. \ref{fig:workflow}. Let's assume that the time of the simulation is $0<t<t_{E}$ and it is subdivided into $n_{TS}$ time steps ranging from $n=1 ... n_{TS}$ and $t_{E}$ is the time at the end of the simulation. At time step $n$ there is no interaction between both code instances, so no contact is detected (Fig. \ref{fig:workflow}a). Then, at time $t^{n+1}$, contact is detected as we have overlapping between both bodies. At this step, the non-penetration boundary conditions are treated kinematically, i.e., as Mulitple Point Constraints (MPC) by the projection of the nodes belonging to the slave surface (deformable's body) to the master surface (rigid's body) using a Dirichlet condition. Then, a local coordinate system with normal-tangent basis vectors $\underline{\boldsymbol{n}}_{j}$ and $\underline{\boldsymbol{t}}_{j}$ is created for each detected node $j$. The contact node is restricted to only move to the tangent line defined by the vector $\underline{\boldsymbol{t}}_{j}$, see Fig. \ref{fig:workflow}e. In a hypothetical frictional contact problem, the friction force would be imposed in this direction as a Neumann boundary condition. In this work, friction is not considered for the low-velocity impact as the relative velocities at the contact zone are sufficiently small. After that, the contact algorithm checks the presence of adhesion or artificial contact nodes, i.e., nodes in traction (Fig. \ref{fig:workflow}c). The reaction contact force $\underline{\boldsymbol{f}}_{j}^{c}$ has to satisfy the following condition:

\begin{equation}\label{eq:anodes}
\underline{\boldsymbol{f}}_{j}^{c} \cdot \underline{\mathbf{n}}_{j} \geq 0
\end{equation}

Those adhesion nodes have to be released, so the current time step $t^{n+1}$ has to be repeated; the $i$ index shown in Fig. \ref{fig:workflow}c and \ref{fig:workflow}d represents the sub-iterations for node release. The whole kinematic constraint process is depicted for two of the contacting nodes in Fig. \ref{fig:workflow}e. The nodes release algorithm for explicit time schemes is described in Algo. \ref{alg:release_explicit} in \ref{sec:algo}. The condition to distinguish a true contact node or an adhesion (artificial) contact node is by means of the contact force (reaction due to the Dirichlet condition). The vector of contact forces using total Lagrangian formulation can be expressed as:

\begin{equation}\label{eq:fcont}
\underline{\boldsymbol{f}}_{j}^{c} = \int_{\Omega_{0}} \underline{\underline{\boldsymbol{B}}}_{0j}^{T} \underline{\underline{\boldsymbol{P}}} \,d\Omega_{0}
\end{equation}
 
\noindent where $\underline{\underline{\boldsymbol{B}}}_{0j}^{T}$ is the matrix containing the derivatives of the shape functions with respect to the reference system and $\underline{\underline{\boldsymbol{P}}}$ is the nominal stress tensor, see \cite{Belytschko2014}.

An exciting aspect of the PDN method is that the computational cost of the projections is very small compared to Penalty or Lagrange approaches \cite{Lapeer2019}. One of the most consuming parts and a vital issue for further research is the contact searching and communication between the subdomains (belonging to different code instances), as stated in the previous work from the authors \cite{Guillamet2022a}. In our case, we use the \verb|PLE++| library \cite{ZavalaPhd2018}, which is an adaptation of the Parallel Location and Exchange \verb|PLE| library \cite{Fournier2014}. The main algorithm of the PDN contact method is described in Algo. \ref{alg:main} in \ref{sec:algo} and the nodes release algorithm for explicit schemes is summarized in Algo. \ref{alg:release_explicit}. The reader is referred to Rivero \cite{Rivero2018PhD} and Guillamet et al. \cite{Guillamet2022a} works for more details on the implementation aspects.

\subsection{Time integration schemes}

{
\newcommand{\vect}[1]{\underline{\boldsymbol{#1}}}
\newcommand{\matr}[1]{\underline{\underline{\boldsymbol{#1}}}}
\newcommand{\displ}{\vect{\textbf{d}}}
\newcommand{\veloc}{\dot{\displ}}
\newcommand{\accel}{\ddot{\displ}}
\newcommand{\mass}{\matr{M}}
\newcommand{\lmass}{\vect{m}}
\newcommand{\fglb}{\vect{f}}
\newcommand{\fint}{\vect{f}^{i}}
\newcommand{\fext}{\vect{f}^{e}}
\newcommand{\fcon}{\vect{f}^{c}}
\newcommand{\fine}{\vect{f}}
\newcommand{\grav}{\vect{g}}

\subsubsection{Deformable body}

Spurious oscillations may appear when using explicit time schemes for dynamic and wave propagation problems such as impact events. These oscillations occur due to the mismatch of two different types of wave components. Thus, dissipative explicit time schemes are often used to reduce the numerical instabilities induced by the spatial and time discretization procedures. Among the many dissipative methods available, the Tchamwa–Wielgosz (TW) explicit scheme \cite{Maheo2009} is beneficial because it damps out the spurious oscillations occurring in the highest frequency domain. This is the time integration scheme selected in this work, but any other explicit time scheme such as the Central Difference (CD) \cite{Belytschko2014} including bulk viscosity could also be used. 

The motion described by the TW scheme is the following:

\begin{equation}
	\veloc^{n+1} = \veloc^{n} + \Delta{t} \accel^{n}
\end{equation}

\begin{equation}
	\displ^{n+1} = \displ^{n} +  \Delta{t} \veloc^{n} + \varphi (\Delta{t})^{2} \accel^{n}
\end{equation}

\noindent where $\displ,~\veloc,~\accel$ are the displacement, velocity and acceleration nodal vectors, respectively; $\Delta{t}$ is the time increment or step size; $n+1$ is the current time step and $n$ is the previous one; $\varphi$ is a numerical viscous parameter, which in the current work is set to $1.033$ \cite{Maheo2009}. The key to the computational efficiency of explicit time integration schemes is the use of the lumped mass matrix for the resolution of the linear system of equations, which is simplified as an easy inversion of the diagonal mass matrix \cite{Belytschko2014}. The global stiffness matrix is not required to be assembled as it is needed for implicit time integration schemes. The explicit time integration scheme solves accelerations, so the discrete momentum equation at time step $n+1$ for a dynamic problem is the following:

\begin{equation}
	\lmass ~ \accel^{n+1} = \fine^{n+1}(\displ^{n+1},t^{n+1}) = \fext(\displ^{n+1},t^{n+1}) - \fint(\displ^{n+1},t^{n+1}) - \fcon(\displ^{n+1},t^{n+1}) 
\end{equation}

\noindent where $\lmass$ is the vector representation of the lumped mass matrix, $\fext$ is the global vector of external forces, $\fint$ is the global vector of the internal forces, and $\fcon$ is the global vector of the contact forces from the Dirichlet condition imposed on that nodes. Thanks to $\lmass$, the acceleration can be computed without invoking any solver as:

\begin{equation}
	\accel^{n+1} = \lmass^{-1} (\fext-\fint-\fcon)
\end{equation}

\subsubsection{Rigid body}

The striker in the present work is considered as a rigid body. The resolution of the equations of motion for the rigid bodies we use a 4th order Runge Kutta scheme. Let’s consider the following differential equation where the right hand side is a function of both time and another function dependent on time.

\begin{equation}
	\frac{dy}{dt} = f(t,y(t))
\end{equation}

From this equation, the Runge-Kutta method estimates the solution at $n+1$ taking into account four evaluations of the right hand side step $dt$ as follows,

\begin{equation}\label{eq:rk4}
\begin{split}
k_{1} = dt \cdot f(t,y(t)) \\ 
k_{2} = dt \cdot f(t+\frac{dt}{2},y(t)+\frac{k_{1}}{2}) \\ 
k_{3} = dt \cdot f(t+\frac{dt}{2},y(t)+\frac{k_{2}}{2}) \\ 
k_{4} = dt \cdot f(t+dt,y(t)+k_{3}) \\ 
y^{n+1}=y(t+dt) = y(t)+\frac{k_{1}}{6}+\frac{k_{2}}{3}+\frac{k_{3}}{4} +\frac{k_{4}}{6} 
\end{split}
\end{equation}

In the present paper, the motion of the striker is solved by making use of the following differential equation:

\begin{equation}
m \accel = m \grav - \fext
\end{equation}

\noindent where $m$ is a scalar value of the mass of the rigid body, $\grav$ is the gravity force vector at the center of mass, $\accel$ is the linear acceleration, and $\fext$ is the external force also at the center of mass from the rigid body. It is worth mentioning that the rigid body is represented by a point (center or mass), so the above vectors have a dimension of 2 for 2-d problems and 3 for 3-d problems. When contact occurs the external force from the rigid body is calculated by $\fext=\sum_{j=1}^{n_{c}}\underline{\boldsymbol{f}}_{j}^{c}$, where $j$ denotes a contact node and $n_{c}$ is the total contact nodes belonging to the deformable body. 

}

\subsection{Mesoscale damage modeling for fiber-reinforced composites}

The mesoscopic length scale is the most suitable for virtual testing of low-velocity impacts on structures made of composite materials. At this scale, the numerical predictions have a good trade-off between information about the damage mechanisms driving the failure process and the structural response without the complexity of dealing with intricate microstructures. It is worth emphasizing that the mesoscopic length scale is not only appropriate for the bottom levels of the building block approach (coupon and elements) \cite{Lopes2009,Tan2015,Furtado2019,Sommer2022} but also for the top levels (sub-components and components) \cite{Reinoso2013,Reinoso2016,Cheng2022}.

From the constitutive modelling viewpoint, the mesoscopic length scale simplifies the intricate microstructure of long fibre composite laminates by homogenising the properties and mechanisms at the lamina level. The outcome is a layered material with two well-defined regions: intralaminar and interlaminar. The former is modelled as a transversally isotropic material, which can fail due to fibre breaking and matrix cracking according to the loading scenario. The latter is modelled as a very thin region, usually tending to zero thickness, where delamination can onset and propagate. 

Regarding the modelling architecture, several strategies exist in the literature suitable for modelling composite at a mesoscopic length scale using FEM \cite{Meer2012}. We adopt a continuous approach for the intralaminar region (Continuum Damage Mechanics (CDM) with linear elements) and a discontinuous one for the interlaminar (Cohezive Zone Model (CZM) with interface elements). The straightforward implementation of this strategy in a standard FEM code aids in preserving the scalability of \verb|Alya| multiphysics \cite{Quintanas2018}. Thus, the mesoscale damage modeling strategy exploits the computational resources to maximize the accuracy of the impacts thanks to very thin meshes.

\subsubsection{Intralaminar damage model}\label{sec:intramodel}

The intralaminar damage model for predicting ply failure is based on the continuum damage mechanics framework. Fiber and matrix cracks are smeared in the continuum and represented by state variables. Accordingly, the crack's kinematics is not explicitly represented, but their effects on the degradation of the capacities of sustaining loads. In turn, the onset and growth of the damage failure mechanisms are governed by the failure surfaces and evolution laws. In this work, we employ a local damage model based on the constitutive modeling framework for long fiber composite materials proposed by Maim\'{i} et al. \cite{Maimi2007b,Maimi2007c}. This framework has been used widely in the literature, demonstrating outstanding accuracy and performance not only for static scenarios \cite{Camanho2007,Bisagni2011,Quintanas2018} but also for impact \cite{Gonzalez2012,Soto2018,Sasikumar2020} and fatigue \cite{Llobet2021a,Llobet2021b}.

The main ingredients of the intralaminar damage model are: i) transversally isotropic elasto-plastic response, ii) damage activation functions related to the different ply failure mechanisms through the maximum strain criterion for the fiber breaking and the LaRC criteria for the matrix cracking, iii) the damage evolution laws are defined to dissipate the fracture energy associated to the opening mode ensuring mesh objectivity by the crack-band theory \cite{Bazant1983}, and iv) the thermodynamic consistency is ensured by imposing irreversibility of the damage variables.

\begin{figure}[htp]
\begin{center}
\includegraphics[width=16.0cm]{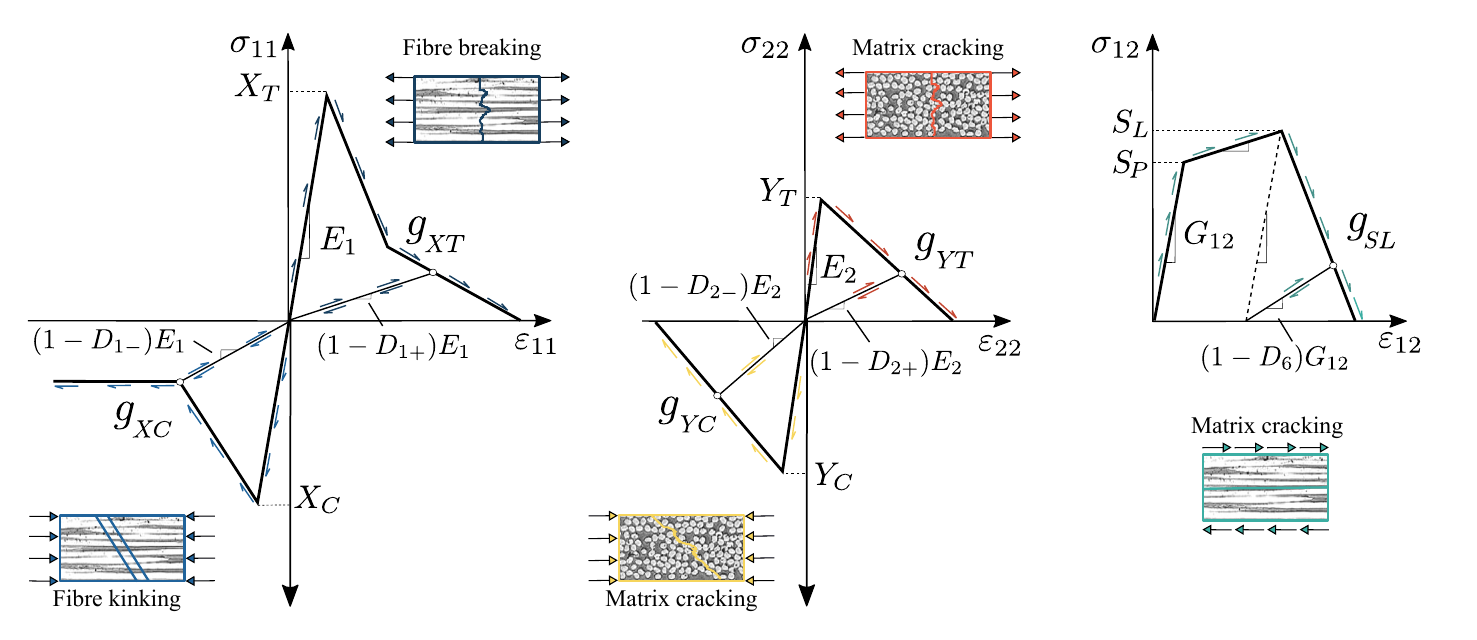} 
\caption{Schematic representation of the intralaminar damage mechanisms. Adapted from \cite{Llobet2021a}.}
\label{fig:dmgintra}
\end{center}
\end{figure}

Fig.~\ref{fig:dmgintra} illustrates the intralaminar failure modes schematically modelled, while Algo.~\ref{alg:cdm} in \ref{sec:algo} summarises the material model workflow. Note that a plastic response under shear loads is considered, and five damage mechanisms are modelled: fibre breaking, fibre kinking, tensile and compressive matrix cracking, and shear matrix cracking. The details of the expressions employed and their justification from a physical standpoint can be found in \cite{Maimi2007b,Maimi2007c,Soto2018}.

Besides the constitutive response, the intralaminar damage model also encloses the computation of the critical time step, which is required by the explicit time integration scheme. For the sake of simplicity, we utilise the same formula of a transversally isotropic material for each element:

\begin{equation}
    \Delta t =  \frac{\ell_c}{c_{D}} = \ell_c \sqrt{\frac{\rho}{\max{C_{ij}}}}
\end{equation}

\noindent where $c_{D}$ is the dilational wavespeed of the material, $C_{ij}$ are the components of the effective stiffness matrix, $\rho$ is the density of the material, and $\ell_c$ is the characteristic element length. Considering a structured hexahedral mesh employed, we approximate the characteristic element length with the element volume $V_{e}$ \cite{Maimi2007c}:

\begin{equation}
	\ell_c \approx \sqrt[3]{V_{e}}
\end{equation}

\subsubsection{Interlaminar damage model}\label{sec:intermodel}

The interlaminar damage model for predicting the onset and propagation of delamination is based on the cohesive zone approach and formulated in the context of damage mechanics. Accordingly, a damage state variable is employed to account for the gradual loss of the bearing capacities of the material in the cohesive zone due to the separation of crack surfaces. In turn, the separation or opening of the crack is represented by a kinematic quantity noted as displacement jump, which is approximated by employing the interface element technology \cite{Allix1996,Ortiz1999,DeBorst2006}. Thus, the interlaminar damage model is a constitutive model that computes the cohesive reactions as a function of displacement jumps. More details of the mesoscale modeling of delamination using cohesive zone models can be found in Carreras et al. \cite{Carreras2021}.

In this work, we use the cohesive zone model proposed by Turon et al. \cite{Turon2006,Turon2018b}, which has been employed extensively in the literature \cite{Guillamet2016,Sarrado2016,Plagianakos2020,Quintanas2020}. The main characteristics of this CZ model are: i) linear response before initiation of the softening, ii) linear relation between the cohesive tractions and crack openings, iii) onset and propagation of the damage in compliance with the Benzeggagh-Kenane criterion, and iv) thermodynamic consistency despite the loading scenario, even when the mix-mode ratio varies. Algo.~\ref{alg:czm} summarizes the cohesive zone model workflow.

Regarding the element technology, we employ zero-thickness interface elements for capturing the delamination, implemented using the formulation presented in \cite{Reinoso2014}. As standard interface elements are used, the integrals are computed using a Newton-Cotes integration scheme to mitigate the spurious oscillations in the traction profile along the interface \cite{Schellekens1993}. The stable time increment, which is necessary for the explicit time integration scheme is obtained through \cite{Soto2018}:

\begin{equation}
  \Delta t_{coh} = \sqrt{\frac{\bar{\rho}}{K_{coh}}}
\end{equation}

\noindent where $\bar{\rho}$ and $K_{coh}$ are numerical parameters known as cohesive surface density and penalty stiffness, respectively. The cohesive surface density for zero-thickness elements is approximated by the expression in \cite{Soto2018b}. In turn, the cohesive penalty stiffness is defined to avoid affecting the compliance of the system as $K_{coh} \ge 50 E_T ⁄ t_{lam}$, where $E_T$ is the transverse elastic modulus and $t_{lam}$ the adjacent laminate thickness \cite{Turon2007}. 

\subsection{Mesh refinement algorithm for interface elements with a cohesive law}\label{sec:mmu}

The mesh multiplication algorithm proposed by Houzeaux et al. \cite{Houzeaux2013} has been extended to deal with the presence of interface elements or even continuum shell element formulations. Focusing on interface elements, they are zero-thickness elements with a cohesive material law that are inserted between plies in a laminated composite material in order to predict the delamination damage mechanism. It is well known that an accurate prediction of the onset and propagation of delamination in composite materials requires very refined meshes, as stated in \cite{Turon2006,Turon2007}. However, depending on the number of interface layers or the geometry size, it can be challenging to place these elements between plies and computationally demanding to solve the problem.

On the other hand, mesh generation of large meshes is often a bottleneck in engineering applications to deal with thousands of millions of elements. Thus, integrated tools for mesh refinement within parallel codes devised for High-Performance systems allow a parallel and fast refinement of the coarse mesh without the need to create the mesh again. 

\begin{figure}[H]
\begin{center}
\includegraphics[width=8.5cm]{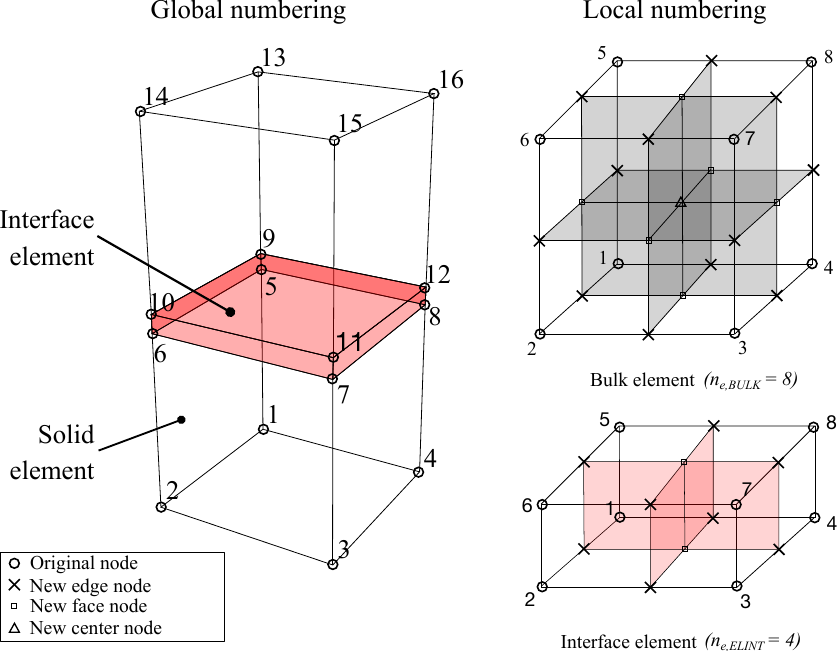} 
\caption{Mesh multiplication between bulk and interface (cohesive) elements.}
\label{fig:mmu}
\end{center}
\end{figure}

Therefore, this paper also introduces a new capability of the mesh multiplication algorithm from \cite{Houzeaux2013}, which enables the refinement of large-scale problems, including interface elements. Let's assume a configuration of two bulk elements together with an interface element between them, as shown in Fig.~\ref{fig:mmu}.

The 8-node interface element ($ELINT$) can only be divided into four elements to avoid the duplication of the element at the interface mid-plane between the bulk elements ($BULK$). The criterion used for the correct division is by making use of the element normal, also known as stacking direction, which is required for the proper behaviour of the element due to its kinematics. Thus, those parallel planes to the element normal are used to divide the element. The dimensions of the new mesh can be calculated as follows:

\begin{equation}\label{eq:mmu}
\scriptsize
\begin{split}
n_{e} = 8\cdot n_{e,BULK}^{0} - n_{e,ELINT}^{0}\cdot4 \\
n_{n} = n_{n}^{0} + n_{edges} + n_{faces} + n_{e,BULK}^{0} - n_{e,ELINT}^{0}-n_{edges,ELINT}^{0}-n_{faces,ELINT}^{0} \\
n_{b} = 4\cdot n_{b}^{0} - 2\cdot n_{b,ELINT}^{0}
\end{split}
\end{equation}

\noindent where $n_{e}$, $n_{n}$ and $n_{b}$ are the total number of elements, nodes and boundaries for the new mesh. In order to refine the hybrid mesh is important to know the total number of $n_{e}^{0}$, $n_{n}^{0}$ and $n_{b}^{0}$ from the original mesh and also information about the edges and faces that have to be divide or not. Algo.~\ref{alg:mmu} summarizes the different steps and functions for the mesh division and reconstruction of the interface domains in a parallel framework.

\section{Benchmark tests}

Three benchmark tests are conducted to validate the application of the parallel partial Dirichlet-Neumann contact algorithm using an explicit time integration scheme. The first example consists of a quasi-static indentation test, which has already been solved using an implicit time integration scheme in \cite{Guillamet2022a}. The solution using explicit analysis is compared with the numerical solution obtained for implicit analysis. The second and the third examples consist of a low-velocity impact event on two coupons manufactured with two well-known material systems for the damage prediction: T800S/M21 and AS4/8552 respectively. Thanks to the proposed algorithm's flexibility and generality, we use a multi-code approach, where the motion of each body (rigid and deformable) is solved using different instances of \verb|Alya|. Regarding the partitioning of the mesh, we use the Space-Filling Curve (SFC) based partitioner described in \cite{Borrell2018}, which performs the partitioning in parallel and maximizes the load balance. It is worth highlighting that all the executions here are in parallel (pre-process, solution, and post-process steps). In all the examples, the contact bodies are discretized with a refined finite element mesh to assess the geometrical localization between both code instances and to obtain an accurate prediction of the contact force. All the simulations are conducted in {\it MareNostrum4} supercomputer. This cluster has 3456 nodes, each of them with 48 processors Intel Xeon Platinum $@$ 2.1 [GHz], giving a total processor count of \num{165888} processors.

\subsection{Quasi-static indentation test}

\begin{figure}[H]
\begin{center}
\includegraphics[width=8.5cm]{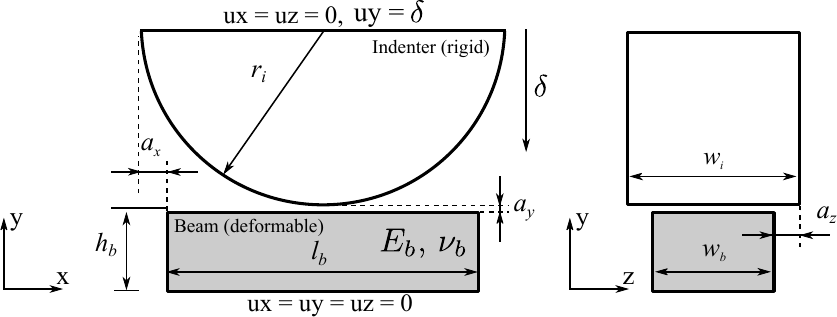} 
\caption{Setup for the quasi-static indentation test. Adapted from \citep{Guillamet2022a}.}
\label{fig:indentation3d-model}
\end{center}
\end{figure}

This example has already been solved using an implicit time solution scheme in \cite{Guillamet2022a,Rivero2018PhD}. In this paper, it is solved as a quasi-static problem using explicit dynamics. The example consists of a rigid rounded head (indenter) and a deformable beam, see Fig.~\ref{fig:indentation3d-model}. The geometrical dimensions of the indenter (rigid body) are $r_i=\,$\SI{1}{\m} and $w_i=\,$\SI{0.5}{\m}, while the beam (deformable body) are $h_b=\,$\SI{0.25}{\m}, $l_b=\,$\SI{1.5}{\m} and $w_b=\,$\SI{0.3}{\m}. The relative position of the indenter with respect to the beam is given by the parameters $a_x=\,$\SI{0.25}{\m}, $a_z=\,$\SI{0.1}{\m} and $a_y=\,$\SI{0.01}{\m} (gap). The beam is modelled with an hyperelastic Neo-Hookean formulation \cite{AnsysMechanical} and finite strains, with material properties $E_b=\,$\textbf{\SI{6.896e+8}{\pascal}} (Young modulus), $\nu_{b}=\,$\SI{0.32}{} (Poisson ratio) and density $\rho=\,$\SI{1000}{\kilogram\per\metre\cubed}. The beam is fully clamped at the bottom face, and a prescribed vertical displacement of $\delta =\,$\SI{0.11}{\m} is applied at the top surface belonging to the indenter. Both bodies are discretized with finite elements using full integration: 8-node linear solid elements for the beam and 4-node linear tetrahedrons for the indenter. The beam has a base mesh of \num{3510} elements, while the indenter has \num{15960} elements. A non-linear dynamic analysis is performed with a total time of the simulation of $\SI{0.05}{\s}$ and a fixed time step of \num{1e-5}. The selected time step value is smaller than the stable time increment, which is \num{2.796e-5}, and no mass scaling is used. In order to perform a quasi-static event and minimize the kinetic energy, a smooth step function (fifth-order polynomial) is applied. This function has the form $A_{0} + (A_{E} - A_{0})\xi^{3}(10 - 15\xi + 6\xi^{2})$ for $t_{0} \leq t < t_{E}$, where $A_{0}$ and $A_{1}$ are the initial and final amplitude, $t_{0}$ and $t_{E}$ are the initial and final time of the simulation and $\xi=\frac{t-t_{0}}{t_{E}-t_{0}}$. This smooth load rate ensures that the first and second time derivatives are zero at the beginning and the end of the transition.

\begin{figure*}[h]
\begin{center}
\includegraphics[width=\textwidth]{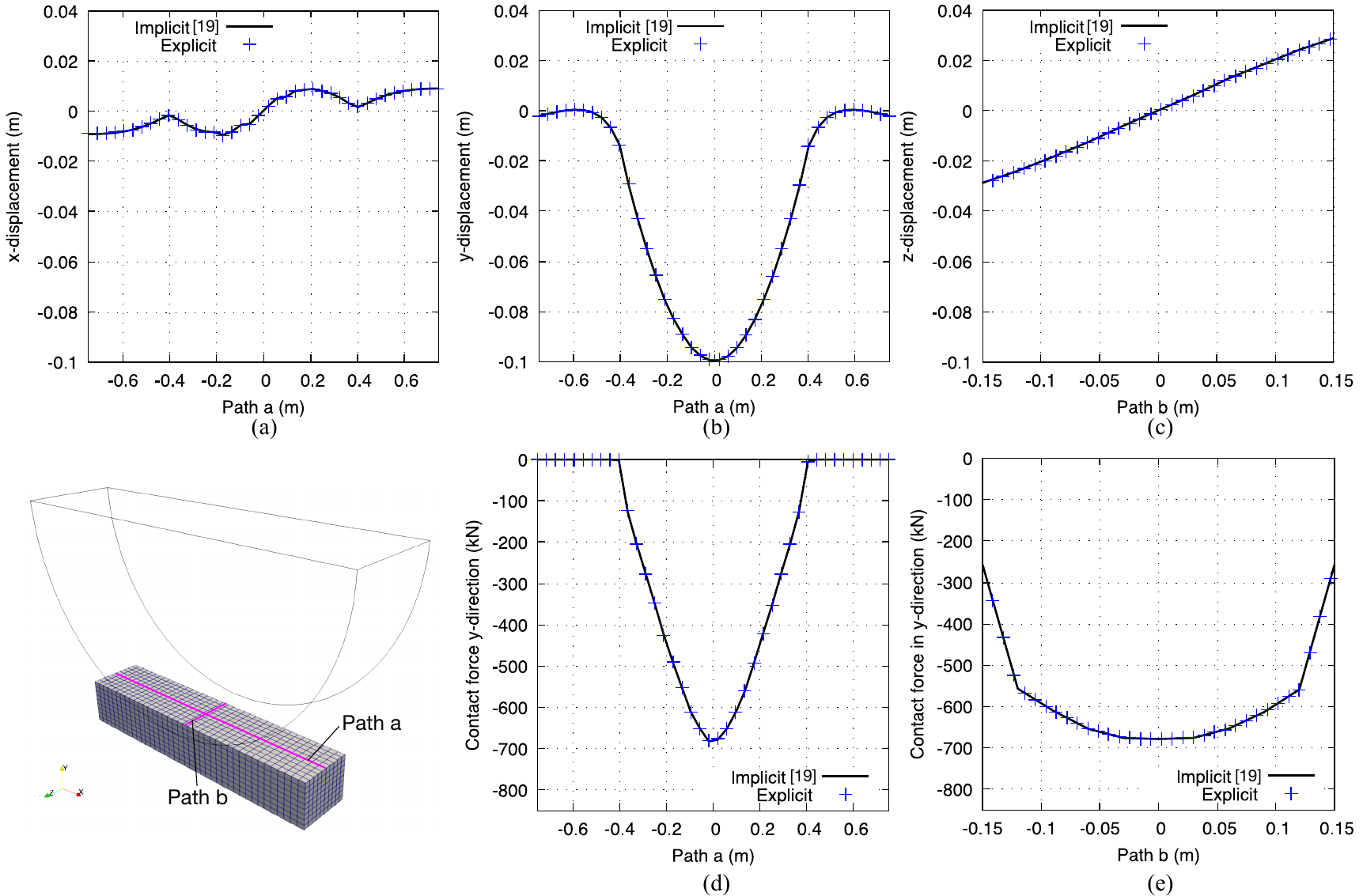} 
\caption{Displacements and contact forces at straight lines path $a$ and $b$. (a) Tangential displacement in x-direction for path $a$. (b) Normal displacement in y-direction for path $a$. (c) Tangential displacement in z-direction for path $b$. (d) Contact force at line path $a$. (e) Contact force at line path $b$.}
\label{fig:indentation3dresults}
\end{center}
\end{figure*}

Displacements and forces obtained at the contact zone are shown in Fig.~\ref{fig:indentation3dresults} for two different paths. Line path $a$ is centered and goes from one side to the other in the length direction of the beam, while line path $b$ is also centered in the width direction. The numerical prediction using the explicit time integration scheme is compared with the implicit solution obtained in \cite{Guillamet2022a}. We can observe an excellent agreement between both numerical predictions in terms of the displacements and contact force.

\subsection{Low velocity impact on a composite plate}\label{sec:lvi}

\begin{figure*}[h]
\begin{center}
\includegraphics[width=\textwidth]{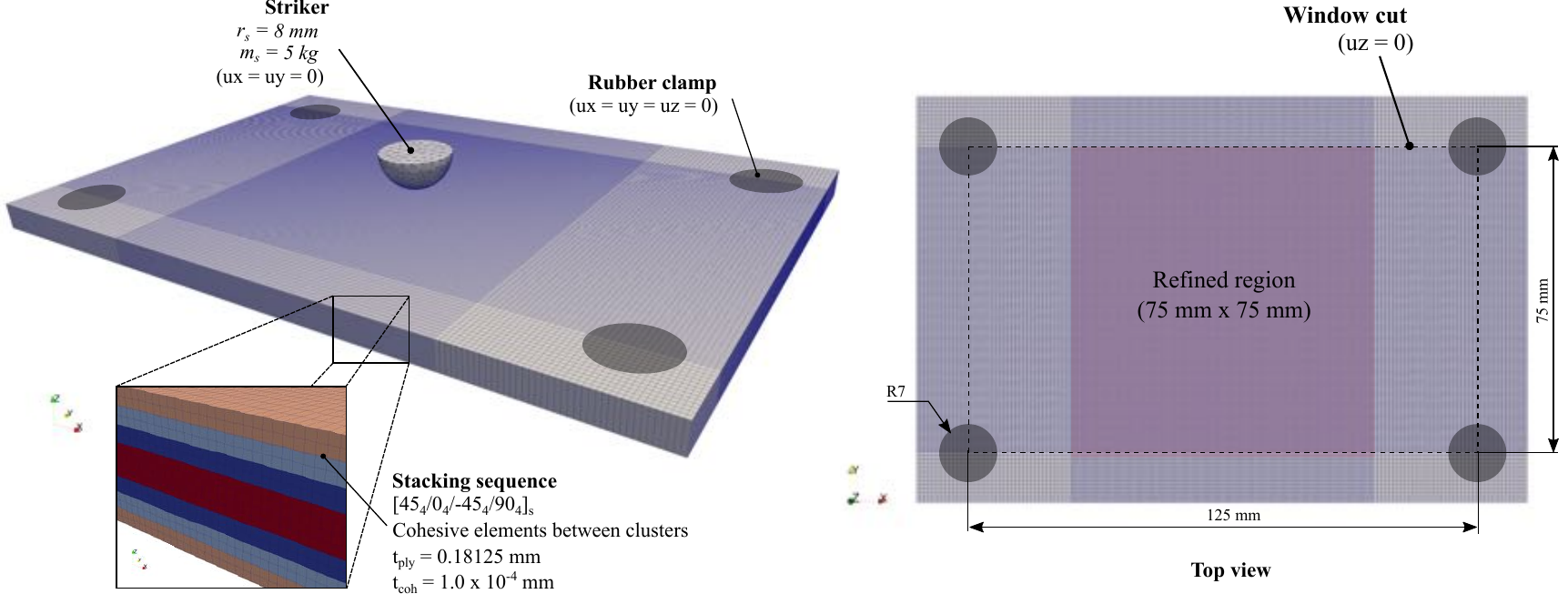} 
\caption{Numerical setup for the low velocity impact test. The layup correspond to the coupon made of AS4/8552 material system. The mesh in this figure corresponds to the based mesh of \num{1472328} elements used for the computational performance analysis.}
\label{fig:lviproblem}
\end{center}
\end{figure*}

The proposed benchmark consists of a drop-weight of a rigid hemispherical striker on a rectangular plate made of composite material, see Fig. \ref{fig:lviproblem}. Two impact scenarios using different material systems, layups, and impact energies are considered for the validation of the proposed framework. The materials selected are the unidirectional prepreg M21/194/34\%/T800S (T800S/M21) and the unidirectional prepreg AS4/8552, both carbon-epoxy systems. On the one hand, the coupon made of T800S/M21 is manufactured by Hellenic Aerospace Industry and tested at Element Materials Technology Seville facilities within the framework of the CleanSky2 SHERLOC project. Most of the material properties from the T800S/M21 are also characterized by Hellenic Aerospace Industry and Element Materials Technology Seville. On the other hand, the coupon made of AS4/8552 is chosen from literature through the works conducted by Gonz\'{a}lez et al. \cite{Gonzalez2012} and Soto et al. \cite{Soto2018}. All the material properties from the aforementioned materials, including damage model parameters, are summarized in Tab. \ref{tab:matprop}. The intralaminar damage model is fed by the in-situ strengths which are calculated following the works by \citet{Furtado2019} and \citet{Soto2018}.

\begin{table}[hbp] 
\centering\scriptsize
\begin{tabular}{lrrrrr}
\hline  
                          &   \multicolumn{3}{c}{T800S/M21}  &  \multicolumn{2}{c}{AS4/8552}  \\
Property                  &     Value         & CV(\%)   & Ref. & Value  &  Ref. \\
\hline 
Density ($t/mm^{3}$)      & \num{1.59e-9}     &   -      & &  \num{1.59e-9} & \cite{Soto2018,Gonzalez2012} \\
\multicolumn{6}{l}{Elastic}  \\
$E_{11}$ (MPa)            &  \num{138.4e+3}   & 1.95    & & \num{128.0e+3} & \cite{Soto2018,Gonzalez2012}\\
$E_{22}=E_{33}$  (MPa)    &  \num{8.54e+3}    &  3      & & \num{7.63e+3} & \cite{Soto2018,Gonzalez2012} \\
$\nu_{12}=\nu_{13}$ (-)   &  \num{0.311}      &  16     & & \num{0.35}    & \cite{Soto2018,Gonzalez2012} \\
$\nu_{23}$  (-)           &  \num{0.45}       &  -      & & \num{0.45}    & \cite{Soto2018,Gonzalez2012} \\
$G_{12}=G_{13}$ (MPa)     &  \num{4.29e+3}    &  3      & & \num{4.358e+3} & \cite{Soto2018,Gonzalez2012}\\
$G_{23}$  (MPa)           &  \num{2.945e+3}   &  -      & & \num{2.631e+3} & \cite{Soto2018,Gonzalez2012} \\
\multicolumn{6}{l}{Strength}  \\
$X_{T}$  (MPa)            & \num{2854.0}      &  4      & & \num{2300.0}    & \cite{Soto2018,Gonzalez2012} \\
$X_{C}$  (MPa)            & \num{1109.0}      &  13     & & \num{1531.0}    & \cite{Soto2018,Gonzalez2012} \\
$Y_{T}$  (MPa)            & \num{56.6}        &  5.8    & & \num{74.2}      &  \\
$Y_{C}$  (MPa)            & \num{250.0}       &         &  \cite{Furtado2019}  & \num{199.8}  & \cite{Soto2018,Gonzalez2012} \\
$S_{L}$  (MPa)            & \num{93.7}        & 0.6     &  & 94.36$^{a}$   & \cite{hex8552report} \\
$\alpha_{o}$ ($^{\circ}$) & \num{53}          &   &  \cite{Maimi2007b} &  53         &   \cite{Maimi2007b}    \\ 
\multicolumn{6}{l}{In-situ strengths$^{c}$}  \\
$Y_{T,int}^{is}$  (MPa)  & \num{132.5} (1$t_{ply}$)      &  - & & \num{117.5} (4$t_{ply}$)  \\
$Y_{T,int}^{is}$  (MPa)  & \num{93.7}  (2$t_{ply}$)      &  - & & \num{117.5} (8$t_{ply}$)  \\
$Y_{T,out}^{is}$  (MPa)  & \num{83.8}  (1$t_{ply}$)      &  - & & \num{74.2}  (4$t_{ply}$)  \\
$Y_{C,int}^{is}$  (MPa)  & \num{250.0} (1$t_{ply}$)      &  - & & \num{199.8} (4$t_{ply}$)  \\
$Y_{C,int}^{is}$  (MPa)  & \num{250.0} (2$t_{ply}$)      &  - & & \num{199.8} (8$t_{ply}$)  \\
$Y_{C,out}^{is}$  (MPa)  & \num{250.0} (1$t_{ply}$)      &  - & & \num{199.8} (4$t_{ply}$)  \\
$S_{L,int}^{is}$  (MPa)  & \num{116.0} (1$t_{ply}$)      &  - & & \num{120.8} (4$t_{ply}$)  \\
$S_{L,int}^{is}$  (MPa)  & \num{116.0} (2$t_{ply}$)      &  - & & \num{120.8} (8$t_{ply}$)  \\
$S_{L,out}^{is}$  (MPa)  & \num{93.7}  (1$t_{ply}$)      &  - & & \num{94.4}  (4$t_{ply}$)  \\
\multicolumn{6}{l}{Fracture toughness}\\
$G_{XT}$  (N/mm)          &  \num{340}        &  & \cite{Furtado2019} &  \num{81.5}  & \cite{Soto2018,Gonzalez2012} \\
$G_{XC}$  (N/mm)          &  \num{60.0}       &  & \cite{Furtado2019} &  \num{106.3} &  \cite{Soto2018,Gonzalez2012} \\
$G_{YT}$  (N/mm)          &  $G_{Ic}$         & 7.3 & &  $G_{Ic}$   & \cite{Soto2018,Gonzalez2012} \\
$G_{YC}$  (N/mm)          &  \num{1.38}$^{b}$ & 20  & & \num{1.313}$^{b}$ & \cite{Soto2018,Gonzalez2012} \\
$G_{SL}$  (N/mm)          &  $G_{IIc}$        & 20  & & $G_{IIc}$  &  \cite{Soto2018,Gonzalez2012}  \\
\multicolumn{6}{l}{Traction separation law}\\
$f_{XT}$  (-)             &  \num{0.1}        & - & & \num{0.1}   & \cite{Soto2018,Gonzalez2012}  \\
$f_{GT}$  (-)             &  \num{0.6}        & - & &  \num{0.6}  & \cite{Soto2018,Gonzalez2012}  \\
$f_{XC}$  (-)             &  \num{0.1}        & - & &  \num{0.1}  &  \cite{Soto2018,Gonzalez2012}  \\
$f_{GC}$  (-)             &  \num{0.9}        & - & &  \num{0.9}  &  \cite{Soto2018,Gonzalez2012}  \\
\multicolumn{6}{l}{Matrix plasticity} \\
$S_{p}$  (N/mm)           &  66.9     & - & \cite{Furtado2019} &  62.0$^{a}$ &   \\
$K_{p}$  (N/mm)           &  0.09     & - & \cite{Furtado2019} &  0.1936$^{a}$ &  \\
\multicolumn{6}{l}{Interface properties} \\
$G_{Ic}$ (N/mm)           & \num{0.308}       & 7.3    & & \num{0.28} &    \cite{Soto2018,Gonzalez2012}  \\
$G_{IIc}$ (N/mm)          & \num{0.828}       & 20     & & \num{0.79} &    \cite{Soto2018,Gonzalez2012}  \\
$\tau_{I}$ (MPa)          & \num{49.2}$^{d}$  & 5.8    & & $Y_{T}$    & \cite{Soto2018,Gonzalez2012}  \\
$\tau_{II}$ (MPa)         & \num{80.7}$^{d}$  & 0.6    & & $S_{L}$    & \cite{Soto2018,Gonzalez2012}  \\
$\eta$ (-)                & \num{1.75}        & -      & & \num{1.45}   &  \cite{Soto2018,Gonzalez2012}  \\
$K_{coh}$ (M/$mm^{3}$)    & \num{1.1e+6}      & -      & & \num{2.5e+4} &  \cite{Soto2018}  \\
\hline
\multicolumn{6}{l}{\tiny $^{a}$ Best fitted based on properties from \cite{hex8552report}} \\
\multicolumn{6}{l}{\tiny $^{b}$ $G_{YC} = G_{SL}/cos(\alpha_{o})$ \cite{Maimi2007b}} \\
\multicolumn{6}{l}{\tiny $^{c}$ Calculated considering plasticity using equations from \cite{Soto2018}} \\
\multicolumn{6}{l}{\tiny $^{d}$ Engineering solution by Turon et al. 2007 \cite{Turon2007} using $N_{e}$=5} \\
\end{tabular}
\caption{Material properties for the M21/194/34\%/T800S (T800S/M21) and Hexply AS4/8552 including damage models parameters.}
\label{tab:matprop}
\end{table}

Both impact case scenarios follow the standard ASTM D7136/D7136M-20 \cite{Astmd7136} for damage resistance evaluation of fiber-reinforced polymers. Each plate has the same dimensions: 150\SI{}{\mm} $\times$ 100\SI{}{\mm} and each of them are supported on a metallic frame with a cut-out of 125\SI{}{\mm} $\times$ 75\SI{}{\mm}. Rubber-tipped clamps fix the plate instance at the four corners. We consider equivalent boundary conditions to represent this experiment. As we can see in Fig.~\ref{fig:lviproblem},  the metallic frame and the rubber clamps from the experiment do not exist as physical entities, so we only consider the contact surface from the rubber cylinder-shaped clamps and the contact edges of the cut-out window of the metallic frame, where we apply the boundary conditions. 

The velocity of the striker is given as an initial condition set in the impact direction, while the remaining degrees of freedom are constrained. The initial velocity of the striker is calculated based on the impact energy of each case study. The initial position of the striker has a gap of \SI{0.01}{\mm} between the striker tip and the top surface of the plate in order to avoid overlapping between bodies at the beginning of the simulation. Moreover, gravity forces are included in both body instances, considering a gravity value of \SI{9.81}{\metre/\second^{2}}. 

We employ 8-node full integration hexahedron elements for the plate using the inter- and intra-laminar damage models described in Sec.~\ref{sec:intermodel} and Sec.~\ref{sec:intramodel} respectively. Cohesive elements are inserted at each interface between different ply angles. It is worth highlighting that other constitutive material models and element technologies would also be feasible in combination with the proposed contact algorithm. With regards to the strikers used for each impact case scenario, they are discretized with 4-node linear tetrahedron elements with a biased mesh of \SI{0.1}{\mm} at the center of the half-sphere and \SI{1}{\mm} at the end of the edge. The total number of elements for the striker used for the T800S/M21 and AS4/8552 materials are \num{32685} and \num{79934}, respectively. Regarding the plates, they both have a refined centered region of 75\SI{}{\mm} $\times$ 75\SI{}{\mm} with an in-plane element size equal or multiple to the ply thickness, depending on the material system in order to guarantee an aspect ratio close or equal to 1.

\subsubsection{Coupon made of T800S/M21 material}
 
This impact coupon has a stacking sequence of $[45/-45/0_{2}/90/0]_{S}$ and is made of T800S/M21. The nominal ply thickness is \SI{0.192}{\mm}. This case study is submitted to an impact energy of \SI{10}{\J}, which falls into the Barely Visible Impact Damage (BVID) analysis. The striker has a diameter of \SI{25}{\mm} and a mass of \SI{2}{\kg}, which is modeled as a rigid body. The global element size for the plate is \SI{1}{\mm}, and each lamina and the clusters of two plies have one element through the thickness. The in-plane element size is \SI{0.192}{\mm} which is equal to the ply thickness resulting in an aspect ratio of 1 for those elements at plies without clustering and located at the refined region. The mesh of the plate has a total of \num{1042525} hexahedron elements ($\approx$ 3.3 million of Degrees Of Freedom (DOF)). The total time for this simulation is set to \SI{5.0}{\ms}. The initial velocity of the striker considering the gap previously mentioned is \SI{3.16}{\m/s}.

\begin{figure}[H]
\begin{center}
\includegraphics[width=\textwidth]{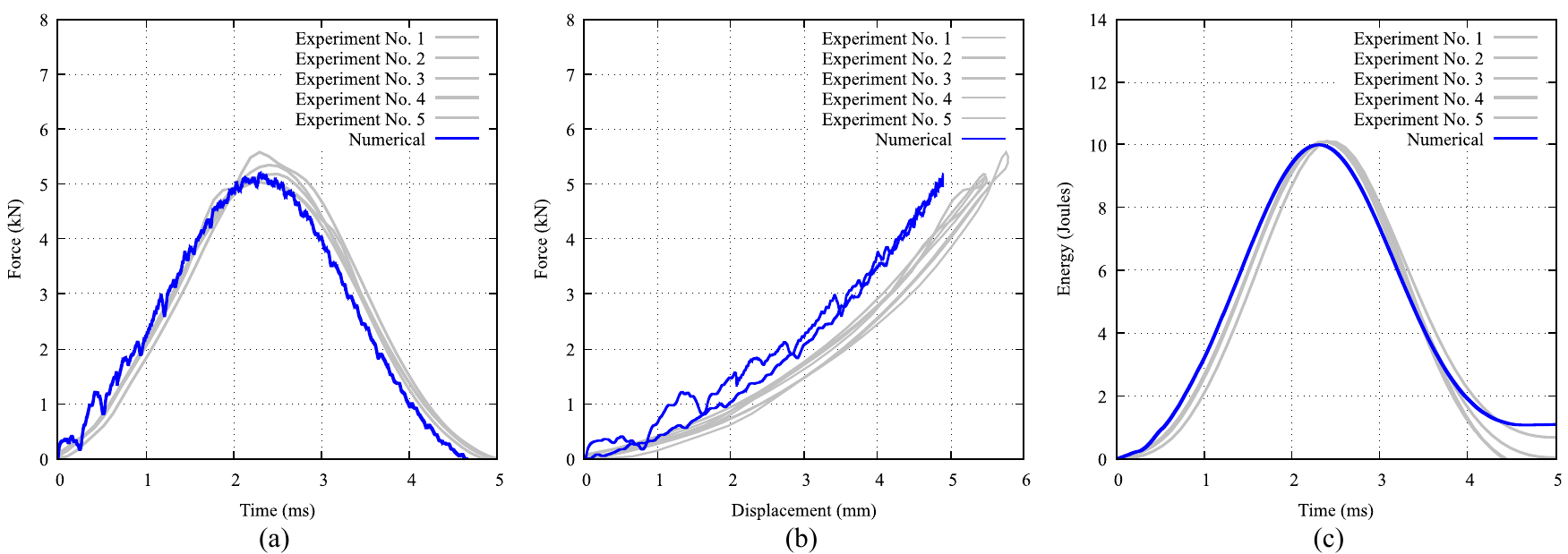} 
\caption{Experimental and numerical curves for the 10J impact on the coupon made of T800S/M21 material. (a) Impact force-time. (b) Impact force-displacement. (c) Energy-time.}
\label{fig:rescoupont800sm21}
\end{center}
\end{figure}

The numerical predictions for this impact case scenario and their comparison with experimental data are shown in Fig. \ref{fig:rescoupont800sm21} and summarized in Tab. \ref{tab:rescoupont800sm21}. The experimental test campaign consisted of testing a batch of five coupons to ensure proper repeatability of the results. The force-time for each impact was recorded with a limited number of points (52 points on average for each impact test). The reduced number of points only allows for validation of the global behavior of the impact case scenario. Energy-time and the force-displacement curves are calculated by integrating once and twice the experimental force history curve. 

As we can see either in Fig. \ref{fig:rescoupont800sm21} and Tab. \ref{tab:rescoupont800sm21} a good agreement is obtained between experiments and numerical predictions. On the one hand, the proposed contact algorithm combined with the proposed damage models is able to capture the maximum impact force very well and the maximum displacement pretty well with errors below 10\%, respectively. On the contrary, the different dissipated energies obtained by the experiments show a high dispersity between them, resulting in difficulty in conducting a fair comparison between the predicted value \SI{1.1}{\J} and the experimental mean value.

\begin{table}[H]
\centering\scriptsize
\begin{tabular}{lrrrr}
\hline  
                                         &  \multicolumn{2}{c}{Experiment} &  Prediction & Difference (\%)   \\
                                         &  Mean                  & Std.   &             &                    \\
\hline
Maximum impact force, $f_{max}^{c}$ (kN) &  5.3    & 0.2  &   5.2          & -1.0   \\
Maximum displacement, $d_{max}$ (mm)     &  5.4    & 0.1  &   4.9          & -9.3 \\
Dissipated energy, $E_{dis}$ (J)         &  0.2    & 0.3  &   1.1          & $>$10   \\ 
\hline
\end{tabular}
\caption{Comparison of numerical results with experimental data for the impact case scenario of the plate made of T800S/M21 material.}
\label{tab:rescoupont800sm21}
\end{table}

\subsubsection{Coupon made of AS4/8552 material}

This second case consists of a coupon made with the AS4/8552 material. The plate has a stacking sequence of $[45_{4}/0_{4}/-45_{4}/90_{4}]_{S}$ with a nominal ply thickness of \SI{0.181}{\mm} resulting a plate thickness of \SI{5.8}{\mm}. This case study has higher energy (\SI{19.3}{\J}) than the previous one, and it also includes clusters of four and eight plies which are potential for extensive matrix cracks and delaminations. The energy of  \SI{19.3}{\J} also falls into BVID analysis. The in-plane element sizes used in \cite{Gonzalez2012} and \cite{Soto2018} are \SI{0.3}{\mm} and \SI{0.5}{\mm} respectively. In the present work, two element sizes are studied using the mesh refinement algorithm described in Sec.~\ref{sec:mmu}, see Tab.~\ref{tab:casestudies}. Only one level of refinement has been applied due to the reduction of the stable time increment and the number of time steps required to finish the simulation. The base mesh for the plate has a total of \num{335622} hexahedron elements ($\approx$ 1 million DOF). The total time for the simulation is set to \SI{5.0}{\ms}. In this case, the striker has a mass of \SI{5}{\kg}, and its radius is \SI{8}{\mm}. The initial velocity of the striker considering the gap previously mentioned is \SI{2.78}{\m/s}.

\begin{table}[H]
\centering\small
\begin{tabular}{rrrrrr}
\hline  
Refinement & Element  & No. element           & No. elem.       & No. nodes      & Initial stable    \\
 level, \textit{ndivi}     & size (mm) & through ply cluster   & plate           & plate          & time increment (s) \\
\hline
 0             &  0.7250    &    1      & \num{335622}   & \num{364320}   & \SI{7.378E-08}{} \\
 1             &  0.3625    &    2      & \num{2109624}  & \num{2219983}  & \SI{3.515E-08}{} \\
\hline
\end{tabular}
\caption{Element sizes used on the coupon made of AS4/8552 material system and initial stable time increment for each case study.}
\label{tab:casestudies}
\end{table}

\begin{figure}[H]
\begin{center}
\includegraphics[width=6.0cm]{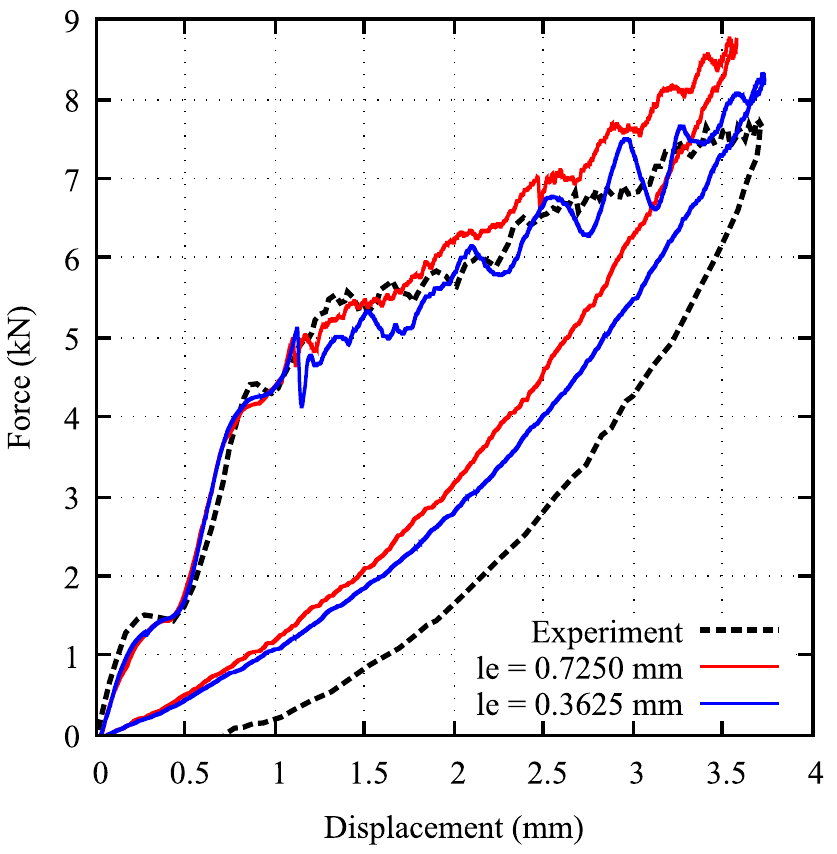} 
\caption{Numerical prediction of the force-displacement curved using two element sizes and correlation with the experiment from Gonz\'{a}lez et al. \cite{Gonzalez2012}. The reader is referred to the web version of this paper for the color representation of this figure.}
\label{fig:rescoupon_f-d}
\end{center}
\end{figure}
\begin{figure}[H]
\begin{center}
\includegraphics[width=6.0cm]{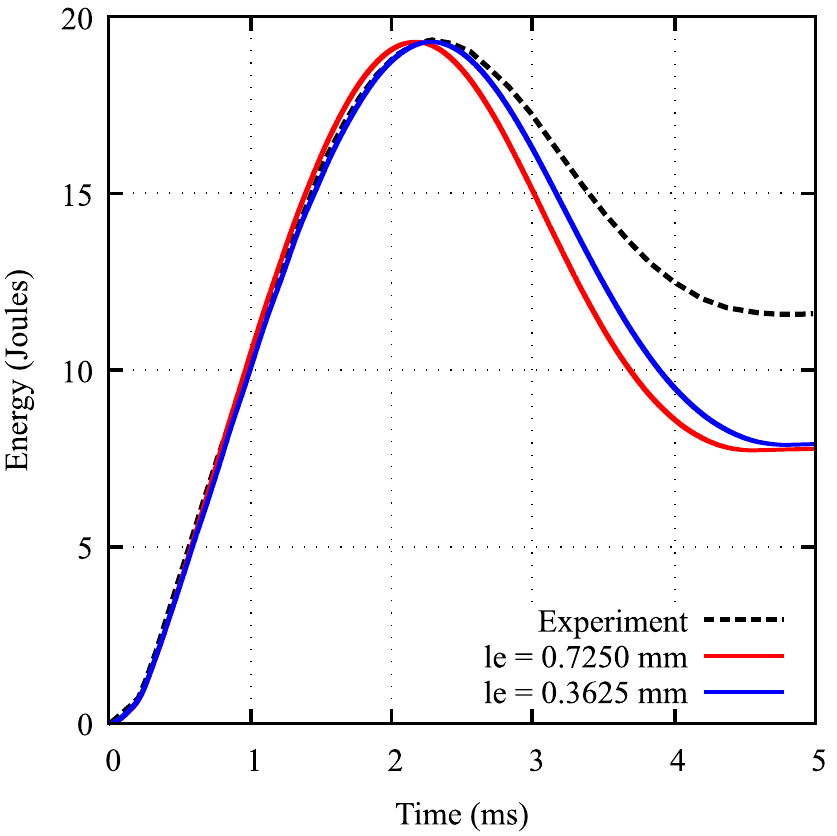} 
\caption{Numerical prediction of the impact energy vs. time using two element sizes and correlation with the experiment from Gonz\'{a}lez et al. \cite{Gonzalez2012}. The reader is referred to the web version of this paper for the color representation of this figure.}
\label{fig:rescoupon_e-t}
\end{center}
\end{figure}

The numerical predictions of the impact force-displacement and energy - time curves are shown in Fig. \ref{fig:rescoupon_f-d} and Fig. \ref{fig:rescoupon_e-t} respectively, using different element sizes. The most important physics variables for a proper validation are summarized in Tab. \ref{tab:rescouponAS48552}. This table compares the experimental results from \cite{Soto2018} with the numerical predictions. 

\begin{table}[H]
\centering\small
\begin{tabular}{lccccc}
\hline  
Case                            & $f_{del}^{c}$ (kN) & $f_{max}^{c}$ (kN) & $d_{max}$ (mm) & $E_{dis}$ (J) & $A_{del}^{proj}$ (mm$^{2}$)     \\
\hline
Experiment \cite{Gonzalez2012}      & 4.41  & 7.74 & 3.72 & 12.03 & 3898.3  \\
Numerical $(l_{e}=\SI{0.7250}{\mm})$    & 4.20  & 8.70 & 3.60 &  7.70 & 4723.1  \\
Numerical $(l_{e}=\SI{0.3625}{\mm})$    & 4.30  & 8.30 & 3.70 &  7.90 & 5249.20 \\
\hline
\end{tabular}
\caption{Comparison of the numerical results obtained with the proposed framework with experimental data from \cite{Soto2018}. $f_{del}^{c}$ is the delamination threshold force, $f_{max}^{c}$ is the maximum contact force, $d_{max}$ is the maximum indentation, $E_{dis}$ is the dissipated energy and $A_{del}^{proj}$ is the projected delamination area.}
\label{tab:rescouponAS48552}
\end{table}

The initial elastic deflection of the plate is very well captured for all the meshes (Fig. \ref{fig:rescoupon_f-d}), meaning that the stiffness of the plate is accurately predicted by the PDN contact algorithm. After that, delamination onset occurs at the top of the elastic part, around \SI{4.5}{\kN}. This point is also very well captured by the interlaminar damage model using cohesive elements between each of the ply clustering. Then, a combination of interlaminar and intralaminar damage occurs until the striker reaches both the maximum load and displacement, resulting with a pretty good prediction as also shown in Tab. \ref{tab:rescouponAS48552}. Since damage appears, the continuum damage models and the characterization of the material properties play a fundamental role in the simulation of this benchmark case. Despite the delamination threshold, maximum force and displacement are very well captured; the dissipated energy and the projected delamination area are overpredicted, see Tab. \ref{tab:rescouponAS48552}. According to Soto et al. \cite{Soto2018}, the projected delamination and the corresponding energy dissipated could be considerably improved when using solid elements with one integration point for the bulk material and cohesive contact surfaces instead of cohesive elements to be able to better predict the delamination shapes at each interface of the layup. This would be a key issue for further research.

\begin{figure}[H]
\begin{center}
\includegraphics[width=\textwidth]{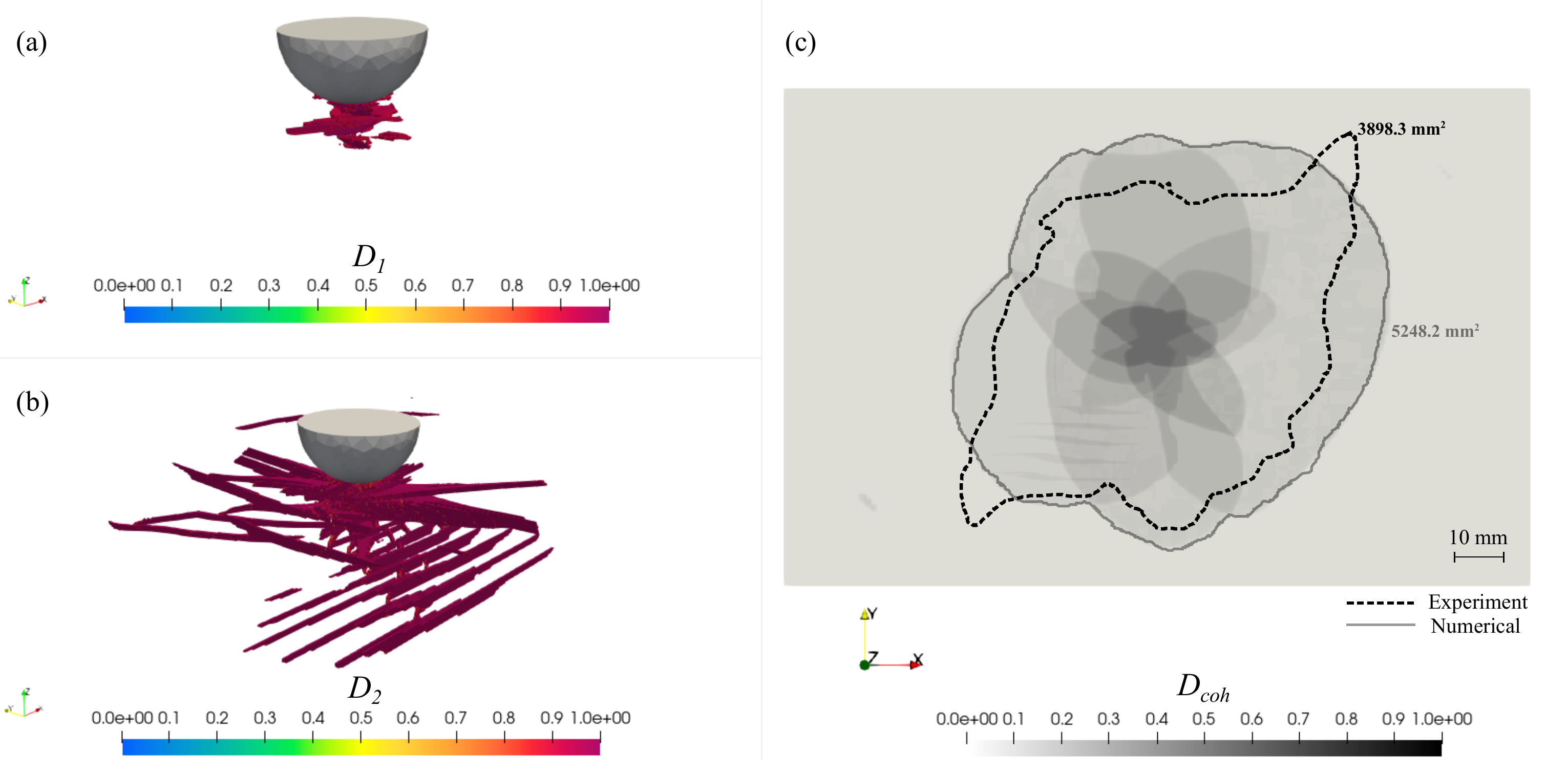} 
\caption{Numerical prediction of the damage occurred in the coupon. (a) Fiber damage, $D_{1}$. (b) Matrix cracking, $D_{2}$. (c) Projected delamination, $D_{coh}$. The numerical result correspond to the most refined mesh. The reader is referred to the web version of this paper for the color representation of this figure.}
\label{fig:rescoupon_dama}
\end{center}
\end{figure}

Fig.~\ref{fig:rescoupon_dama} depicts and aims to quantify the most important failure mechanisms that appear on the plate. Fiber damage is represented by damage variable $D_{1}$, which includes both fiber breakage and fiber kinking, see Fig.~\ref{fig:rescoupon_dama}a. As we can see, this source of damage is not the most predominant and mostly appears at the bottom of the striker. Matrix cracking is represented with the damage variable $D_{2}$, which includes matrix tension and compression (Fig. \ref{fig:rescoupon_dama}b). Finally, the last source of damage is delamination (Fig. \ref{fig:rescoupon_dama}b). Its prediction is compared with the shape obtained from the experiment, which is represented in dashed lines. As we discussed previously, this source of damage is overpredicted for all the element sizes studied, see Tab. \ref{tab:rescouponAS48552} and further research would be required in that direction as the values of the material properties, and the damage models play a fundamental role. Furthermore, the extensive matrix cracks and delamination predicted for this impact case scenario corroborate the experimental observations by Gonz\'{a}lez et al. \cite{Gonzalez2011} on the effect of ply clustering to originate extensive matrix cracks and large delaminations.

\subsection{Parallel performance}

\begin{figure}[h]
\begin{center}
\includegraphics[width=8.5cm]{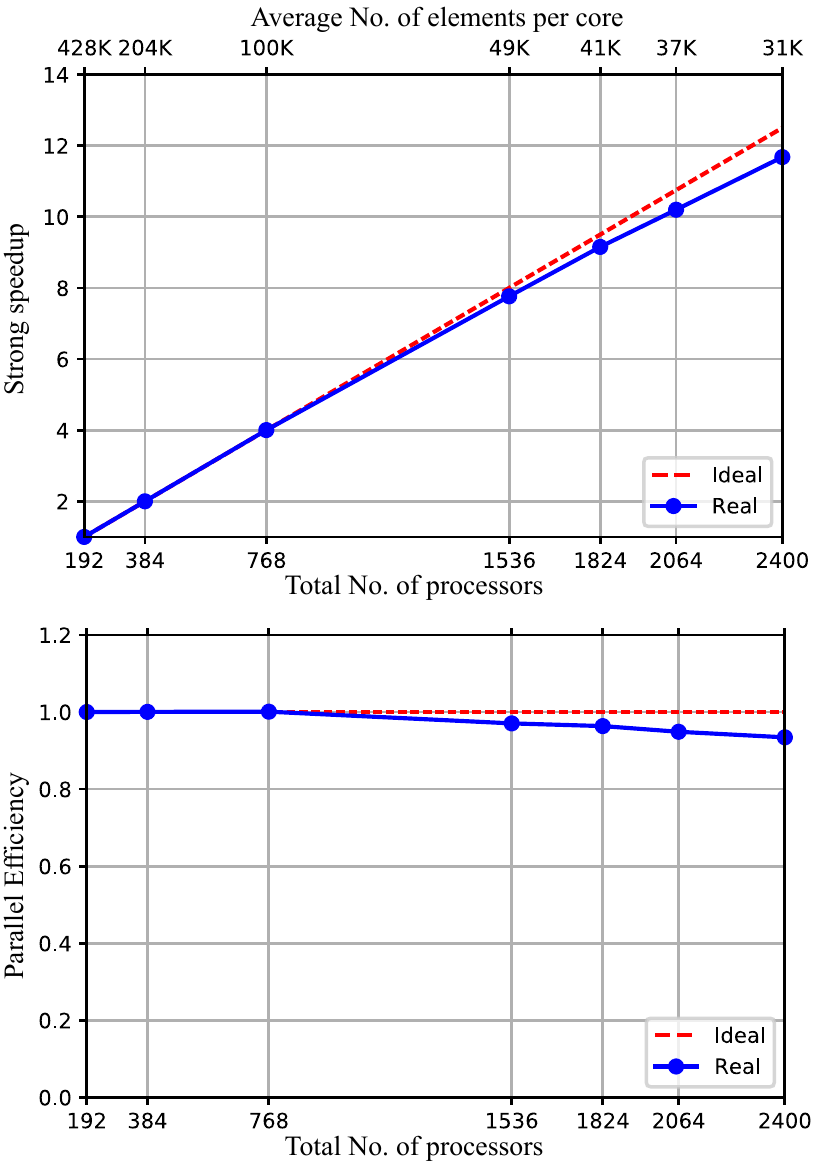} 
\caption{Strong scalability of the low velocity impact test with a plate mesh of 74M hexahedron elements. The model corresponds to the benchmark case using AS4/8552 material system.}
\label{fig:scal74M}
\end{center}
\end{figure}

The speedup and the parallel efficiency of the proposed contact algorithm for solving low-velocity impact events are evaluated in this section. The main objective of this computational analysis is to show the efficiency of the parallel algorithm for explicit schemes rather than the resulting non-linear behavior produced by the damage models used in this case. All the executions are conducted in {\it MareNostrum4} supercomputer. A strong scalability analysis has been conducted using a larger mesh than the ones studied in Sec.~\ref{sec:lvi}. The model corresponds to the AS4/8552 impact case scenario. The new mesh has a total of 74M elements with 228M of DOF, which results from a base mesh of \num{1472328} elements using two levels of the mesh refinement algorithm. The refined mesh region has an element size of 0.362\SI{}{\mm} $\times$ 0.362\SI{}{\mm} $\times$ 0.362\SI{}{\mm} and a global in-plane element size of \SI{1}{\mm}. Strong scalability consists of fixing the mesh and solving the problem with a different number of processors, Central Processing Unit (CPU). The strong speedup is calculated as $\frac{t_{0}}{t_{N}}$ while the parallel efficiency is calculated as $\frac{t_{0}N_{0}}{t_{N}N}$, where $N$ is the number of processors and $t_{0}$ is the reference simulation time for $N_{0}$ processors. The number of processors used for this analysis ranges from 192 to \num{2400}. Due to the resolution of the problem following a multibody/multicode approach, the number of processors for the striker is fixed to 16 (sufficiently for its mesh) while the number of processors for the plate is changed. It is worth mentioning that the strong computational effort falls in the resolution (deformation) of the plate and the localization and exchange of information phases, as explained in \cite{Guillamet2022a}. Due to the small time step in this simulation, \SI{9.261E-09}{s}, the simulations for the scalability curve are limited to the first \num{7460} time steps. The end of the execution (last time step) corresponds to an impact force of approximately \SI{1}{kN}, which falls into the linear elastic regime of the force-displacement curve shown in Fig.~\ref{fig:rescoupon_f-d}. The strong speedup and parallel efficiency are shown in Fig.~\ref{fig:scal74M}. The ideal scalability and efficiency are represented with a dashed line.

The results obtained in Fig.~\ref{fig:scal74M} show that the scalability of the problem in explicit analysis is really good up to \num{2400} processors using a mesh of 74M elements. The parallel efficiency is maintained above 90\%, which demonstrates the good scalability of the proposed framework to deal with large-scale problems. This linear behavior is also shown in Tab. \ref{tab:cputimes} where we summarize the total CPU time for each execution using a different number of processors while maintaing fixed the size of the problem. 

\begin{table}[H]
\centering\scriptsize
\begin{tabular}{ccccccc}
\hline  
   \multicolumn{7}{c}{No. of CPUs} \\
192     &   384 & 768 & 1536 & 1824 & 2064  & 2400 \\
\hline
17:20   &  08:33 & 4:15 & 02:11 & 01:51 & 01:39 & 01:27 \\
\hline
\end{tabular}
\caption{Total CPU time expressed in hh:mm for different executions of the low-velocity impact simulation considering a fixed mesh of 74M of elements (228M of DOF) with a total of 7460 time steps. This CPU time includes the preprocess, where two mesh refinement levels are performed and the solution of the contact problem within the elastic regime of the force-displacement curve. All the executions use double precision and they are executed using Intel\textsuperscript{\tiny\textregistered} compiler and MPI libraries version 2017.4.}
\label{tab:cputimes}
\end{table}

It is also worth mentioning that the application of the proposed contact algorithm in explicit dynamics improves both the speedup and the parallel efficiency in comparison to an implicit resolution for the deformable body (plate), as already studied by the authors in \cite{Guillamet2022a}. This improvement in computational performance is mainly attributed to the time integration scheme for the deformable body. In explicit dynamics, it is not required to invert the global matrix of the system. In this case, the unknown is the acceleration, and the system is solved directly using the lumped mass matrix and the global force vector on the right-hand side. The reader is referred to \cite{Belytschko2014} for more details.

\section{Conclusions}

In this paper, we apply the parallel PDN contact algorithm to simulate low-velocity impact events on fiber-reinforced polymer composites using a High-Performance Computing environment. Existing damage models from the literature have been implemented in our multiphysics finite element code \verb|Alya| to simulate the material damage. Moreover, we introduce a new capability in the in-house mesh refinement algorithm to deal with cohesive elements and other element types, such as continuum shell elements. This is really attractive as we can refine the finite element mesh at the beginning of the simulation with a meager computational cost.

We validate the whole framework with several benchmark tests. The last example corresponds to a well-known low-velocity impact test following the ASTM standard for damage resistance analysis. In this case, we study two impact case scenarios with two different material systems: the T800S/M21 and the AS4/8552, obtaining excellent predictions for impact behavior and pretty good damage occurrence compared to experimental data from the literature. Additionally, the mesh refinement algorithm's capabilities have been demonstrated for the plate made of AS4/8552 material.

Finally, we evaluate the parallel performance of the impact simulation. Despite not using "very" large meshes for the physics validation cases, we have generated a new larger mesh using the mesh refinement algorithm. The reason behind this is the stable time increment, which becomes smaller as the element size decrease. The new mesh has 74M hexahedron elements (228M of DOF) using full integration. An excellent computational efficiency (above 90\%) has been obtained up to 2400 CPUs, demonstrating its applicability to solve large mesh models ranging from micro-scale to macro-scale.

\section{Challenges and future work}

We have demonstrated the potential application of the parallel PDN contact algorithm for low-velocity impact events and its parallel efficiency for large models compared to traditional Penalty or Lagrange contact-based methods in HPC systems. As we previously commented, we use full integration elements for all the examples. The use of reduced integration elements, which are more appropriate for explicit schemes and overcome the well-known locking pathologies from solid brick elements, would considerably increase the speedup of the simulations.

A key aspect of research is applying the proposed method to deal with cohesive contact surfaces. In such cases, we should use a deformable-to-deformable contact (bilateral approach) to solve the contact behavior between the adjacent materials. One of the main ingredients would be adequate conservative interpolations between both contact surfaces.

In the previous examples, we do not consider friction as it plays a minor role in the overall behavior of the proposed contact problems. The use of friction forces on the PDN contact method has already been studied for simple cases in the Ph.D. thesis from M. Rivero \cite{Rivero2018PhD}; however, further research would be required, particularly for 3-d problems.

Last but not least, the localization of contact nodes and the communication between subdomains created by the domain decomposition method is a crucial issue for further research as it is the main bottleneck regarding the computational efficiency of contact algorithms.

%
%
%

\section*{Acknowledgements}

This work has received funding from the Clean Sky 2 Joint Undertaking (JU) under grant agreements No. 807083 and No. 945521 (SHERLOC project). The JU receives support from the European Union’s Horizon 2020 research and innovation program and the Clean Sky 2 JU members other than the Union. The authors gratefully acknowledge Hellenic Aerospace Industry for manufacturing of the coupons made of T800S/M21 material and Kirsa Mu\~{n}oz and Miguel \'{A}ngel Jim\'{e}nez from Element Materials Technology Seville for conducting the experimental impact tests and providing all the experimental data. A. Quintanas-Corominas acknowledges financial support from the European Union-NextGenerationEU and the Ministry of Universities and Recovery, Transformation and Resilience Plan of the Spanish Government through a call of the University of Girona (grant REQ2021-A-30). G. Guillamet thankfully acknowledges the computer resources at \textit{MareNostrum} and the technical support provided by Barcelona Supercomputing Center (IM-2021-2-0022). Last but not least, the authors would also like to thank the late Claudio Lopes for all the interesting discussions and contributions to the simulation of impact events and damage on composites.

\appendix
\section{Algorithms}
\label{sec:algo}

Here we summarize the main algorithms of the whole modeling framework to solve low-velocity impact events for damage resistance of fiber-reinforced polymer composites by making use of High-Performance Computing.

\begin{algorithm}[H]
\scriptsize
\caption{Main code for the partial Dirichlet-Neumann (PDN) contact algorithm.}
\label{alg:main}
\begin{algorithmic}[1]
\item[~]{This PDN contact algorithm is treated as a coupling problem between two or more body instances. In the present algorithm, we describe the contact algorithm between two code instances: a rigid body represented by the domain $\Omega_{a}$ and the deformable body represented by the domain $\Omega_{b}$. The coupling is performed through the exchange of boundary conditions at the contact interface following a Gauss-Seidel strategy. At each time step, contact detection is done for both instances, and synchronization and localization is executed. When contact is detected (at least one boundary node belonging to the deformable body is penetrated inside the rigid body), the rigid one computes and sends to the deformable body all the information required for the enforcement of the kinematic boundary conditions. The reader is referred to the Ph.D. from Rivero \cite{Rivero2018PhD}, or \cite{Guillamet2022a} for more details on the implementation aspects of the proposed contact algorithm.}
\Require $\Omega_{a}$, $\Omega_{b}$
\Loop~time
\State Compute time step, $t^{n+1}$
  \Loop~reset
   \If{Rigid body, $\Omega_{a}$}                 
     \State Contact detection (localization)                  \Comment{Contact detection \& localization, Algo.~1~in \cite{Guillamet2022a}}
     \State Exchange data: receive $\underline{\mathbf{f}}^{c}$ from $\Omega_{b}$  \Comment{Exchange \& communication data, Algo.~2~in \cite{Guillamet2022a}}
     \State \verb|call|~\emph{calculateProjections()}               \Comment{Projections \& local coordinate system, Algo.~3~in \cite{Guillamet2022a}}
     \State \verb|call|~\emph{RK4Scheme()}                    \Comment{Solve system}
     \State Exchange data: send projection data to $\Omega_{b}$               \Comment{Exchange \& communication data, Algo.~2~in \cite{Guillamet2022a}} \label{lin:conta11}
   \EndIf
    \If{Deformable body, $\Omega_{b}$}
      \State Contact detection (localization)                    \Comment{Contact detection \& localization, Algo.~1~in \cite{Guillamet2022a}}
      \State Exchange data: receive data (projections) from $\Omega_{a}$  \Comment{Algo.~2~in \cite{Guillamet2022a}}
      \State \verb|call|~\emph{EssentialBoundaryCondition()}              \Comment{Contact nodes \& Dirichlet condition, Algo.~4~in \cite{Guillamet2022a}}
      \State \verb|call|~\emph{ExplicitScheme()}                    \Comment{Solve system}
      \State \verb|call|~\emph{ReleaseNodes()}                      \Comment{Algo.~\ref{alg:release_explicit}}
      \State Exchange: send $\underline{\mathbf{f}}^{c}$ to $\Omega_{a}$ \Comment{Exchange \& communication data, Algo.~2~in \cite{Guillamet2022a}}    \label{lin:conta1}
   \EndIf
   \If{$\emph{kfl\_reset} = 0$}  
       \State \verb|exit| loop reset 
   \EndIf
   \EndLoop
\EndLoop
\end{algorithmic}
\end{algorithm}

\begin{algorithm}[H]
\scriptsize
\caption{ \textit{ReleaseNodes()} algorithm for explicit time integration schemes}
\label{alg:release_explicit}
\begin{algorithmic}[1]
\item[~]{This algorithm is executed concurrently and for each subdomain at the end of the time step $t^{n+1}$. The $\emph{kfl\_reset}$ is the key flag for the repetition of the current time step $t^{n+1}$ when exists adhesion contact nodes. The sign of the contact force is checked according to Eq.~\ref{eq:anodes}. The key flag to release the adhesion nodes is called \emph{kfl\_nodes\_to\_release}. Then the adhesion contact nodes are released (as free non-contacting nodes), and the time step is repeated, activating the reset key flag. As all the subdomains require to know if the time step has to be repeated or not, the \verb|MPI_MAX| is in charge to collect the value of the reset for all the subdomains of the mesh.}
    \State $\emph{kfl\_reset} \gets 0$                                     \label{lin:rel1}
      \State Get contact force $\underline{\mathbf{f}}^{c}$ and mark adhesion nodes \label{lin:rel2}
      \If{\emph{kfl\_nodes\_to\_release}}                                 \label{lin:rel3}
        \State Adhesion nodes are set to free nodes                       \label{lin:rel4}
        \State $\emph{kfl\_reset} \leftarrow 1$                         \label{lin:rel5}
      \EndIf                                                            \label{lin:rel6}
    \State \verb|call|~\verb|MPI_MAX|$( \emph{kfl\_reset} )$                 \label{lin:rel8}
\end{algorithmic}
\end{algorithm}

\begin{algorithm}[H]
\scriptsize
\algrenewcommand\algorithmicrequire{\textbf{Input}}
\algrenewcommand\algorithmicensure{\textbf{Output}}
\caption{Recursive mesh multiplication algorithm}
\label{alg:mmu}
\begin{algorithmic}[1]
\item[~]{The level of mesh refinement is set with the parameter $ndivi$. $n_{e}$, $n_{n}$ and $n_{b}$ are the total number of elements, nodes and boundaries of the new mesh. The same parameters with the superscript $0$ indicate the initial dimension of the mesh. In order to define the dimensions of the new mesh is necessary to know the total number of edges ($n_{edgg}$) and faces $n_{facg}$ of the initial mesh. Then, once the dimensions are known, the  \emph{DivideMesh()} subroutine is in charge of doing the following actions: i) divide each edge and face from the initial mesh, ii) define the new element connectivities, iv) define the new element boundary connectivities and v) assign the material codes and the corresponding fields such as material coordinate systems. The last step of the mesh division algorithm is to reconstruct the interface domains through the \emph{ReconstructInterfaceDomains()} subroutine. The reader is referred to Houzeuax et al. \cite{Houzeaux2013} for more details on the implementation and parallel aspects of the proposed algorithm.}
\Require $n_{e}^{0}$, $n_{n}^{0}$, $n_{b}^{0}$ , $ndivi$
\Ensure $n_{e}$, $n_{n}$, $n_{b}$
\For{$idivi = 1,ndivi$ } \label{lin:mmu0}
	\State $n_{edgg} = $ \emph{GetEdges()} \label{lin:mmu1}
	\State $n_{facg} = $ \emph{GetFaces()} \label{lin:mmu2}
	\State $n_{e},n_{n},n_{b} = $ \emph{GetDimensions()} \label{lin:mmu3} \Comment{Eq.~\ref{eq:mmu}}
	\State \verb|call|~\emph{DivideMesh()} \label{lin:mmu4}               \Comment{Sec.~3.2 \cite{Houzeaux2013}}
	\State \verb|call|~\emph{ReconstructInterfaceDomains()} \label{lin:mmu5} \Comment{Sec.~3.2 \cite{Houzeaux2013}}
\EndFor 
\end{algorithmic}
\end{algorithm}

\newcommand{\vect}[1]{\underline{\boldsymbol{#1}}}
\newcommand{\matr}[1]{\underline{\underline{\boldsymbol{#1}}}}

\begin{algorithm}[H]
\scriptsize
\algrenewcommand\algorithmicrequire{\textbf{Input}}
\algrenewcommand\algorithmicensure{\textbf{Output}}
\caption{Workflow of the intralaminar damage model.}
\label{alg:cdm}
\begin{algorithmic}[1]
\item[~]{The strain and stress tensors, $\vect{\varepsilon}$ and $\vect{\sigma}$, are defined in the material coordinate system using compact notation \cite{Belytschko2014}. The superscripts ${n}$ and ${n+1}$ define the past and current time steps, respectively. The subscripts $N$ indicate the four damage mechanisms associated with the loading function $\phi_{N}$ and internal threshold variables $r_{N}$ (fibre breaking, fibre kinking, tensile matrix cracking, and compressive matrix cracking). In turn, the subscript $M$ indicates the five uniaxial damage states $D_{M}$, represented in \ref{fig:dmgintra}. The required material properties are i) elastic properties ($E_{11}, E_{22}, \nu_{12}, \nu_{23}, G_{12}$ ), ii) ply strengths ($X_{T}, X_{C}, Y_{T}, Y_{C}, S_{L}$), iii) fracture toughness ($G_{XT}, G_{XC}, G_{YT}, G_{YC}, G_{SL}$) associated with the damage mechanism, and iv) yield strength and hardening ($S_{p}, K_{p}$); all these properties can be obtained through standardised tests or computational micromechanics simulations \cite{Lopes2016}. The required parameters are: characteristic element length $\ell_c$ \cite{Maimi2007c} and state variables at the past time step, i.e. $\vect{\varepsilon}^{n}_{p}$, and $r^{t}_{M}$. At the initial time step, the state variables are initialised as $\vect{\varepsilon}^{n}_{p} = 0$ and $r^{n}_{M}=1$.}
\Require $\vect{\varepsilon}^{n+1}$, $\vect{\varepsilon}_{p}^{n}$, $r_{M}^{n}$, $\ell_c$, \emph{material properties}
\Ensure $\vect{\sigma}^{n+1}$, $\vect{\varepsilon}_{p}^{n+1}$, $r_{M}^{n+1}$
\State $\vect{\varepsilon}_{p}^{n+1}(\vect{\varepsilon}_{p}^{n})$ \Comment{Plastic strains, yield function~in \cite{Soto2018}}
\State $\vect{\varepsilon}_{e}^{n+1} \leftarrow \vect{\varepsilon}^{n+1} - \vect{\varepsilon}_{p}^{n+1}$ \Comment{Effective elastic strains}
\State $\vect{\sigma}_{e}^{n+1} \leftarrow \matr{H}^{-1} \cdot \vect{\varepsilon}_{e}^{n+1}$ \Comment{Effective compliance matrix $\matr{H}$~in \cite{Lopes2016}}g
\State $\phi_{M}^{n+1}(\vect{\sigma}_{e}^{n+1})$ \Comment{Loading functions (failure criteria), Eqs.~8, 13, 20, 21~in \cite{Maimi2007b}}
\State $r_{N}^{n+1}(\phi_{N}^{n+1},r_{N}^{n})$ \Comment{Damage thresholds, Eqs.~24, 26~in \cite{Maimi2007b}}
\State $D_{M}^{n+1}(r_{N}^{n+1})$ \Comment{Damage state variables according \cite{Soto2018} and Eq~6~in \cite{Maimi2007b}}
\State $\vect{\sigma}^{n+1} \leftarrow \matr{H}^{-1}(D_{M}^{n+1}) \cdot \vect{\varepsilon}_{e}^{n+1}$ \Comment{Nominal compliance matrix $\matr{H}(D_{M}^{n+1})$~in \cite{Lopes2016}}
\end{algorithmic}
\end{algorithm}

\newcommand{\dcohe}{D} 
\newcommand{\jump}{\vect{\Delta}}
\newcommand{\tcohe}{\vect{\tau}}

\begin{algorithm}[H]
\scriptsize
\algrenewcommand\algorithmicrequire{\textbf{Input}}
\algrenewcommand\algorithmicensure{\textbf{Output}}
\caption{Workflow of the cohesive zone model.}
\label{alg:czm}
\begin{algorithmic}[1]
\item[~]{The displacement jumps and interface tractions, $\jump = \{\Delta_{1},\Delta_{2},\Delta_{3}\}^{T}$ and $\tcohe = \{\tau_{1},\tau_{2},\tau_{3}\}^{T}$, are defined at the mid-plane being $1$ and $2$ tangential and $3$ normal directions. The superscripts ${t}$ and ${t+1}$ define the past and current time steps, respectively. In turn, the subscript $M$ indicates the pure-mode I and II openings associated with the opening directions, $\text{I} \leftrightarrow \{3\}$ and $\text{II} \leftrightarrow {\{1,2\}}$. The latter is also referred with the subscript $sh$ in \cite{Turon2018b}. The required input parameters are i) onset displacement jumps ($\Delta_{Mo}$), ii) critical displacement jumps ($\Delta_{Mc}$), iii) penalty stiffness ($K_{M}$), and iv) Benzeggagh-Kenane exponent for the mixed-mode ratio ($\eta$). The onset and critical jumps can be obtained from the cohesive strengths ($\tau_{M}$) and fracture toughness material properties by $\Delta_{Mo} = \tau_{M} / K_{M}$ and $\Delta_{Mc} = 2 G_{M} / \tau_{M}$, respectively. The damage threshold state variable at the past time $r_{\dcohe}^{n}$, which is initialised at the initial time step as $r_{\dcohe}^{n}=0$, is also required to evaluate the model.}
\Require $\jump^{n+1}$, $r_{\dcohe}^{n}$, \emph{material properties}
\Ensure $\tcohe^{n+1}, r_{\dcohe}^{n+1}$
\State $K_{B}^{n+1}(\jump^{n+1})$ \Comment{Local mixed-mode penatly stiffness, Eq.~13~in \cite{Turon2018b}}
\State $B^{n+1}(\jump^{n+1})$ \Comment{Local mixed-mode ratio, Eq.~17~in \cite{Turon2018b}}
\State $\lambda_{o}^{n+1}(B^{n+1},K_{B}^{n+1})$ \Comment{Local mixed-mode onset jump, Eq.~26~in \cite{Turon2018b}}
\State $\lambda_{c}^{n+1}(B^{n+1},K_{B}^{n+1},\lambda_{o}^{n+1})$ \Comment{Local mixed-mode propagation jump, Eq.~24~in \cite{Turon2018b}}
\State $\lambda^{t+1}(\jump^{t+1})$ \Comment{Local mixed-mode equivalent jump, Eq.~12~in \cite{Turon2018b}}
\State $H^{n+1}(\lambda^{n+1},\lambda_{o}^{n+1},\lambda_{c}^{n+1})$ \Comment{Loading function (failure criteria), Eq.~20~in \cite{Turon2018b}}
\State $r_{\dcohe}^{n+1}(H^{n+1},r_{\dcohe}^{n})$ \Comment{Damage threshold, Eq.~21~in \cite{Turon2018b}}
\State $\dcohe^{n+1}(r_{\dcohe}^{n+1},\lambda_{o}^{n+1},\lambda_{c}^{n+1})$ \Comment{Damage state, Eq.~20~in \cite{Turon2018b}}
\State $\tcohe_{coh}^{n+1}(\dcohe^{n+1},\jump^{t+1})$ \Comment{Cohesive tractions, Eq.~7~in \cite{Turon2018b}}
\State $\tcohe_{con}^{n+1}(\jump^{n+1})$ \Comment{Contact tractions, Eq.~8~in \cite{Turon2018b}}
\State $\tcohe^{n+1} \leftarrow  \tcohe_{coh}^{n+1} + \tcohe_{con}^{n+1}$
\end{algorithmic}
\end{algorithm}

\bibliographystyle{elsarticle-harv}
\bibliography{Guillamet2022b.bib}

\end{document}